\documentclass[prb,aps,twocolumn,amsmath,amssymb]{revtex4}
\usepackage{graphicx,bm}
\usepackage{dcolumn}

\begin{document}

\title{$\bm{3/2}$-Body Correlations and  Coherence in Bose-Einstein Condensates}

\author{Takafumi Kita}
\affiliation{Department of Physics, Hokkaido University, Sapporo 060-0810, Japan}

\begin{abstract}
We construct a variational wave function for the ground state of weakly interacting bosons that gives
a lower energy than the mean-field Girardeau-Arnowitt (or Hartree-Fock-Bogoliubov) theory.
This improvement is brought about by incorporating the dynamical $3/2$-body processes 
where one of two colliding non-condensed particles drops into the condensate and vice versa.
The processes are also shown to transform the one-particle excitation spectrum
into a bubbling mode with a finite lifetime even in the long-wavelength limit.
These $3/2$-body processes, which give rise to dynamical exchange of particles between the non-condensate reservoir and condensate absent in ideal gases,
are identified as a key mechanism for realizing and sustaining macroscopic coherence in Bose-Einstein condensates.
\end{abstract}

\maketitle

\section{Introduction}

Among the fundamental problems in the theory of Bose-Einstein condensation (BEC) are to clarify (i) how the interaction between 
particles changes the properties of the condensate and one-particle excitations from those of ideal gases,\cite{Huang87,GB68,ZUK77,Politzer96,TK16}
and (ii) how the macroscopic coherence indispensable for superfluidity emerges.
This paper makes a contribution to these issues by constructing a variational wave function for the ground state
with a new ingredient, i.e.,  the $3/2$-body processes where a collision of two non-condensed particles 
throws one of them into the condensate and vice versa.
These are dynamical processes beyond the scope of the mean-field treatment that have not been considered non-perturbatively.
This wave function is given as a {\em superposition in terms of the number of condensed particles} within the fixed-number formalism, 
instead of the total number of particles in a subsystem as discussed by Anderson,\cite{Anderson66}
where depleted particles serve as the particle reservoir for the condensate exchanging particles dynamically.
Thus, the superposition, which is indispensable for bringing macroscopic coherence to the condensate,\cite{Anderson66}
emerges naturally due to the interaction
and is also maintained dynamically.
The $3/2$-body processes are also shown to transform the free-particle spectrum of non-condensed particles in ideal gases into that of a bubbling mode
with an intrinsic decay rate, as expected naturally in the presence of the dynamical exchange of particles between the non-condensate reservoir and condensate.

\begin{figure}[b]
  \begin{center}
   \includegraphics[width=0.9\linewidth]{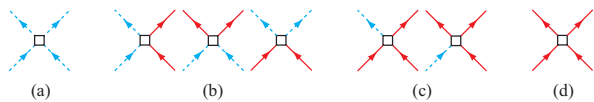}
  \end{center}
  \caption{ Classification of collision processes in homogeneous Bose-Einstein condensates according to the number of non-condensed particles involved. 
  A broken (full) line denotes the condensate (a non-condensed particle), and a square represents the symmetrized interaction vertex.\cite{AGD63}}
  \label{Fig1}
\end{figure}

In 1959, Girardeau and Arnowitt\cite{GA59} constructed a variational wave function for the ground state of homogeneous weakly interacting bosons
so that it is the vacuum of Bogoliubov's quasiparticle operators.\cite{Bogoliubov47}
They used it to evaluate the ground-state energy incorporating two-body interactions of non-condensed particles,
i.e., process (d) in Fig.\ \ref{Fig1}, in addition to processes (a) and (b) of the Bogoliubov theory.\cite{Bogoliubov47}
However, this apparent improvement
brought about an unphysical energy gap in the one-particle excitation spectrum,\cite{GA59,Griffin96}
unlike the Bogoliubov spectrum with a gapless linear dispersion and an infinite lifetime in the long-wavelength limit,\cite{Bogoliubov47} 
which is in contradiction to the Hugenholtz-Pines theorem\cite{HP59} or Goldstone's theorem I.\cite{Goldstone61,GSW62,Weinberg96}
This fact suggests that something crucial may be missing from the Girardeau-Arnowitt wave function, which still remains unidentified explicitly.

The key observation here is that process (c) in Fig.\ \ref{Fig1}, which involves a smaller number of non-condensed particles than (d),
makes no contribution to the energy in the Girardeau-Arnowitt theory.\cite{GA59}
Hence, improving the variational state so as to make process (c) active is expected to lower the energy further
and approximate the true ground state more closely.
Such a state will be constructed below.

It is also interesting to see how process (c) affects properties of one-particle excitations.
We investigate this using the moment method developed previously.\cite{TKK16}
Widely accepted results on the excitations may be summarized as follows:
(i) a finite repulsive interaction between particles turns the free one-particle spectrum $\propto k^2$ of ideal gases 
into the Bogoliubov spectrum $\propto k$ with a lifetime $\tau\propto k^{-5}$ that tends to infinity as the wavenumber $k$ approaches $0$;\cite{Bogoliubov47,Beliaev58,GN64,SK74,WG74,Griffin93,PS02,PS03,OKSD05,Leggett06}
(ii) the Bogoliubov mode is also identical to the density-fluctuation mode (phonons) in the two-particle channel;\cite{GN64,SK74,WG74,Griffin93,PS02,OKSD05,Leggett06}
(iii) the Bogoliubov mode is the Nambu-Goldstone mode of broken $U(1)$ symmetry.\cite{Griffin93,PS03,WM13}
On the other hand, an alternative picture has been presented recently based on a self-consistent perturbation expansion satisfying Goldstone's theorem I and conservation laws 
simultaneously:\cite{Kita09,Kita10,Kita14}
(i$'$) the excitation in the one-particle channel is a bubbling mode with a finite lifetime $\tau<\infty$ even for $k\rightarrow 0$;\cite{Kita11,TK14,TKK16}
(ii$'$) excitations in the one- and two-particle channels are different from each other;\cite{Kita10,Kita11,TKK16}
(iii$'$) the distinct modes in the two channels correspond to two different proofs of Goldstone's theorem,\cite{GSW62,Weinberg96}
the contents of which are not equivalent and should be distinguished as I and II.
Indeed, it has been shown\cite{Kita11,Kita14} that the first proof using the invariance of the effective action under a linear transformation\cite{GSW62,Weinberg96}
is relevant to the poles of the one-particle Green's function in the context of BEC,
whereas the second (and more familiar) one based on the vacuum expectation of the commutator of the current and field is concerned with the poles of the two-particle Green's function.

Here, it will be shown that including process (c) in Fig.\ \ref{Fig1} naturally produces the dynamical exchange of particles between the non-condensate reservoir and condensate, thereby giving rise to an intrinsic decay rate for non-condensed particles even for $k\rightarrow 0$.
Thus, the result here also supports (i$'$), as does our previous study.\cite{TKK16} The finite lifetime of non-condensed particles
may be regarded as a crucial element for realizing and sustaining temporal coherence in the condensate.

This paper is organized as follows. 
Section \ref{sec:formulation} constructs a variational wave function
for the ground state  with the $3/2$-body processes, 
obtains an expression for the ground-state energy, and derives the equations to determine the energy minimum.
Section \ref{sec:NR} presents numerical results for the ground-state energy, one-particle 
excitation spectrum, and superposition over the number of condensed particles obtained using our wave function.
Section \ref{sec:summary} summarizes the paper. 
Appendix A outlines how to obtain an expression for the energy functional
for the Girardeau-Arnowitt wave function. 
Appendix B gives a detailed derivation of the energy functional for our variational wave function with the $3/2$-body processes.
Appendix C describes how to calculate the moments of the one-particle excitation spectrum.

\section{Formulation\label{sec:formulation}}
\subsection{System}

We consider a system of ${\cal N}$ identical particles with mass $m$ and spin 0 in a box of volume ${\cal V}$ described by the Hamiltonian\cite{AGD63,Kita-Text}
\begin{align}
\hat H\equiv &\, \sum_{\bf k}\varepsilon_k \hat c_{{\bf k}}^\dagger\hat c_{{\bf k}}+\frac{1}{2{\cal V}}\sum_{{\bf k}{\bf k}'{\bf q}}U_q
\hat c_{{\bf k}+{\bf q}}^\dagger\hat c_{{\bf k}'-{\bf q}}^\dagger\hat c_{{\bf k}'}\hat c_{{\bf k}}  .
\label{H}
\end{align}
Here $\varepsilon_k\equiv\hbar^2 k^2/2m$ is the kinetic energy,
$(\hat c_{{\bf k}}^\dagger,\hat c_{{\bf k}})$ are the field operators satisfying the Bose commutation relations, and $U_q$ is the interaction potential.
We aim to describe the ground state of Eq.\ (\ref{H}) involving BEC in the ${\bf k}={\bf 0}$ state.
It is convenient for this purpose to classify $\hat H$ according to the number of non-condensed states involved as
\begin{align}
\hat H=\hat H_0+\hat H_1+\hat H_{3/2}+\hat H_{2} .
\label{H-decomp}
\end{align}
Each contribution on the right-hand side is given in terms of the primed sum $\displaystyle{\sum_{\bf k}}'\equiv \sum_{\bf k}(1-\delta_{{\bf k}{\bf 0}})$ as
\begin{subequations}
\label{H-decomp2}
\begin{align}
\hat H_0\equiv &\, \frac{1}{2{\cal V}}U_{0}\hat c_{\bf 0}^\dagger\hat c_{\bf 0}^\dagger\hat c_{\bf 0}\hat c_{\bf 0},
\label{H_0}
\\
\hat H_{1}\equiv &\, {\sum_{{\bf k}}}'  \varepsilon_k\hat c_{{\bf k}}^\dagger\hat c_{{\bf k}}
+\frac{1}{{\cal V}}{\sum_{{\bf k}}}' \left(U_{0}+U_{k}\right)\hat c_{\bf 0}^\dagger\hat c_{\bf 0}\hat c_{{\bf k}}^\dagger\hat c_{{\bf k}}
\notag \\
&\,
+\frac{1}{2{\cal V}}{\sum_{{\bf k}}}'U_{k}\left(\hat c_{\bf 0}^\dagger\hat c_{\bf 0}^\dagger\hat c_{{\bf k}}\hat c_{-{\bf k}}
+\hat c_{-{\bf k}}^\dagger\hat c_{{\bf k}}^\dagger\hat c_{\bf 0}\hat c_{\bf 0} \right),
\label{H_1}
\\
\hat H_{3/2}\equiv&\,\frac{1}{{\cal V}} {\sum_{{\bf k}_1{\bf k}_2}}' U_{k_1}\left(\hat c_{\bf 0}^\dagger\hat c_{{\bf k}_1+{\bf k}_2}^\dagger\hat c_{{\bf k}_2}\hat c_{{\bf k}_1}
+\hat c_{{\bf k}_1}^\dagger\hat c_{{\bf k}_2}^\dagger\hat c_{{\bf k}_1+{\bf k}_2}\hat c_{\bf 0}\right),
\label{H_3/2}
\\
\hat H_{2}\equiv&\, \frac{1}{2{\cal V}}{\sum_{{\bf k}{\bf k}'{\bf q}}}'U_q
\hat c_{{\bf k}+{\bf q}}^\dagger\hat c_{{\bf k}'-{\bf q}}^\dagger\hat c_{{\bf k}'}\hat c_{{\bf k}} .
\label{H_2}
\end{align}
\end{subequations}
The interactions in Eqs.\ (\ref{H_0})-(\ref{H_2}) are expressible diagrammatically as (a)-(d) in Fig.\ \ref{Fig1}, respectively,
by symmetrizing the interaction potential as $U_{k_1}\rightarrow(U_{k_1}+U_{k_2})/2$ in Eq.\ (\ref{H_3/2}) and 
$U_{q}\rightarrow(U_{q}+U_{|{\bf k}+{\bf q}-{\bf k}'|})/2$ in Eq.\ (\ref{H_2}).

\subsection{Number-conserving operators}

Following Girardeau and Arnowitt,\cite{GA59,Girardeau98} we introduce the number-conserving creation-annihilation operators
as follows.
First, orthonormal basis functions for the one-particle state ${\bf k}={\bf 0}$ are given by
\begin{align}
|n \rangle_{\bf 0}\equiv \frac{(\hat c_{\bf 0}^\dagger)^{n}}{\sqrt{n!}}  |0\rangle\hspace{5mm}(n=0,1,2,\cdots,{\cal N}),
\label{|n>_0-def}
\end{align}
where $|0\rangle$ is the vacuum defined by (i) $\langle 0|0\rangle=1$ and (ii) $\hat c_{\bf k}|0\rangle=0$ for any ${\bf k}$.\cite{Kita-Text}
The ground state without interaction  is $|{\cal N} \rangle_{\bf 0}$.
Second, we introduce operators $(\hat\beta_{\bf 0}^\dagger,\hat\beta_{\bf 0})$ by\cite{GA59,Girardeau98}
\begin{align}
\hat\beta_{\bf 0}^\dagger|n \rangle_{\bf 0}\equiv  |n+1 \rangle_{\bf 0},
\hspace{5mm}
\hat\beta_{\bf 0} | n\rangle_{\bf 0} \equiv \left\{\begin{array}{ll}\vspace{1mm} | n-1\rangle_{\bf 0} & :n\geq 1 \\ 0 & :n=0\end{array}\right. .
\label{beta_0-def}
\end{align}
These operators are also expressible in terms of $(\hat c_{\bf 0}^\dagger,\hat c_{\bf 0})$ as $\hat\beta_{\bf 0}^\dagger= \hat c_{\bf 0}^\dagger (1+\hat c_{\bf 0}^\dagger\hat c_{\bf 0})^{-1/2}$ 
and $\hat\beta_{\bf 0}=  (1+\hat c_{\bf 0}^\dagger\hat c_{\bf 0})^{-1/2}\hat c_{\bf 0}$.\cite{Girardeau98}
We then define the number-conserving creation-annihilation operators for ${\bf k}\neq {\bf 0}$ by
\begin{align}
\hat{\tilde c}_{{\bf k}}^\dagger\equiv \hat c_{{\bf k}}^\dagger \hat\beta_{\bf 0} ,\hspace{5mm} \hat{\tilde c}_{{\bf k}}\equiv \hat c_{{\bf k}} \hat\beta_{\bf 0}^\dagger.
\label{tc-def}
\end{align}
Operator $\hat{\tilde c}_{{\bf k}}^\dagger$  has the physical meaning of exciting a particle from the condensate to the state ${\bf k}\neq {\bf 0}$. 

It follows from Eq.\ (\ref{beta_0-def}) that 
\begin{align}
\hat\beta_{\bf 0}^\nu(\hat\beta_{\bf 0}^\dagger)^\nu |n \rangle_{\bf 0}=|n \rangle_{\bf 0},\hspace{5mm}(\hat\beta_{\bf 0}^\dagger)^\nu \hat\beta_{\bf 0}^\nu |n \rangle_{\bf 0}=
\left\{
\begin{array}{ll} |n \rangle_{\bf 0}  & : \nu\leq n 
\\
0 & : \nu> n \end{array}\right. 
\label{beta-beta^dagger}
\end{align}
holds for $\nu=1,2,\cdots$; thus, $\hat\beta_{\bf 0}^\nu(\hat\beta_{\bf 0}^\dagger)^\nu=1$ and $(\hat\beta_{\bf 0}^\dagger)^\nu \hat\beta_{\bf 0}^\nu\!\approx\! 1$. 
For $\nu\ll {\cal N}$, the latter approximation becomes practically exact in the weak-coupling region where the ground state is composed of  
the kets $|n\rangle_{\bf 0}$ with $n=O({\cal N})$.
This fact also implies that the operators in Eq.\ (\ref{tc-def}) satisfy the commutation relations of bosons almost exactly in the weak-coupling region as 
\begin{align}
\bigl[\hat{\tilde c}_{{\bf k}},\hat{\tilde c}_{{\bf k}'}^\dagger\bigr]\approx \delta_{{\bf k}{\bf k}'}, \hspace{5mm}\bigl[\hat{\tilde c}_{{\bf k}},\hat{\tilde c}_{{\bf k}'}\bigr]=0,
\label{tc-commute}
\end{align}
where $[\hat A,\hat B]\equiv\hat A\hat B-\hat B\hat A$. 
Hereafter we replace ``$\approx$'' in Eq.\ (\ref{tc-commute}) by ``$=$''.

\subsection{Girardeau-Arnowitt wave function}

Next, we introduce the Girardeau-Arnowitt wave function in a form different from the original one\cite{GA59} for our convenience.
We define a pair operator $\hat \pi^\dagger$ with non-condensed states by
\begin{subequations}
\begin{align}
\hat\pi^\dagger\equiv \frac{1}{2}{\sum_{{\bf k}}}'\phi_{{\bf k}}\hat c_{{\bf k}}^\dagger\hat c_{-{\bf k}}^\dagger,
\label{pi-def}
\end{align}
where  $\phi_{{\bf k}}$ is a variational parameter with $\phi_{-{\bf k}}=\phi_{{\bf k}}$ by definition.
Its number-conserving correspondent $\hat{\tilde\pi}^\dagger$ is given by
\begin{align}
\hat{\tilde\pi}^\dagger\equiv \hat\pi^\dagger\hat\beta_{\bf 0}^2 = \frac{1}{2}{\sum_{{\bf k}}}'\phi_{{\bf k}}\hat{\tilde c}_{{\bf k}}^\dagger\hat{\tilde c}_{-{\bf k}}^\dagger ,
\label{tpi-def}
\end{align}
\end{subequations}
satisfying
\begin{align}
\bigl[\hat{\tilde c}_{{\bf k}},\hat{\tilde\pi}^\dagger\bigr]=\phi_{{\bf k}}\hat{\tilde c}_{-{\bf k}}^\dagger .
\label{[tc,tpi]}
\end{align}
Using them, we can express the Girardeau-Arnowitt wave function as
\begin{align}
|\Phi_{\rm GA}\rangle\equiv &\,{\cal A}_{\rm GA}\exp\left({\hat{\tilde\pi}^\dagger}\right) |{\cal N}\rangle_{\bf 0}
\notag \\ 
=&\, {\cal A}_{\rm GA}\sum_{\nu=0}^{[{\cal N}/2]}\frac{(\hat\pi^\dagger)^{\nu}}{\nu!}
|{\cal N}-2\nu\rangle_{\bf 0} ,
\label{|Phi_GA>}
\end{align}
where  $[{\cal N}/2]$ denotes the largest integer that does not
exceed ${\cal N}/2$, and ${\cal A}_{\rm GA}$ is the normalization constant determined by $\langle\Phi_{\rm GA}|\Phi_{\rm GA}\rangle=1$.

The ket of Eq.\ (\ref{|Phi_GA>}) is characterized by
\begin{align}
\hat{\tilde\gamma}_{{\bf k}} |\Phi_{\rm GA}\rangle=0\hspace{5mm}({\bf k}\neq{\bf 0}),
\label{gamma|Phi_GA>=0}
\end{align}
i.e., $|\Phi_{\rm GA}\rangle$ is the vacuum of the number-conserving quasiparticle operator
\begin{align}
\hat{\tilde\gamma}_{{\bf k}}\equiv u_{{\bf k}}\hat{\tilde c}_{{\bf k}}-v_{{\bf k}}\hat{\tilde c}_{-{\bf k}}^\dagger ,
\label{gamma-def}
\end{align}
where $u_{{\bf k}}$ and $v_{{\bf k}}$ are defined by
\begin{align}
u_{{\bf k}}\equiv \frac{1}{(1-|\phi_{{\bf k}}|^2)^{1/2}} ,
\hspace{5mm}
v_{{\bf k}}\equiv \frac{\phi_{{\bf k}}}{(1-|\phi_{{\bf k}}|^2)^{1/2}},
\label{uv-def}
\end{align}
satisfying $u_{-{\bf k}}=u_{{\bf k}}=u_{{\bf k}}^*$, $v_{-{\bf k}}=v_{{\bf k}}$, and
 $u_{{\bf k}}^2-|v_{{\bf k}}|^2=1$.
To prove Eq.\ (\ref{gamma|Phi_GA>=0}),
let us operate $\hat{\tilde c}_{{\bf k}}$ on Eq.\ (\ref{|Phi_GA>}) and transform the resulting expression as
\begin{align}\hat{\tilde c}_{{\bf k}} |\Phi_{\rm GA}\rangle=&\,{\cal A}_{\rm GA}\bigl[\hat{\tilde c}_{{\bf k}},e^{\hat{\tilde\pi}^\dagger}\bigr]|{\cal N}\rangle_{\bf 0}
={\cal A}_{\rm GA}\bigl[\hat{\tilde c}_{{\bf k}},\hat{\tilde\pi}^\dagger\bigr]e^{\hat{\tilde\pi}^\dagger}|{\cal N}\rangle_{\bf 0}
\notag \\
=&\,\phi_{{\bf k}}\hat{\tilde c}_{-{\bf k}}^\dagger |\Phi_{\rm GA}\rangle
,\notag
\end{align}
where we used $\hat{\tilde c}_{{\bf k}}|{\cal N}\rangle_{\bf 0}=0$ and Eq.\ (\ref{[tc,tpi]}).
Multiplying the equation in terms of $|\Phi_{\rm GA}\rangle$ by $u_{{\bf k}}$, we obtain Eq.\ (\ref{gamma|Phi_GA>=0}).

It follows from Eqs.\ (\ref{tc-commute}) and (\ref{uv-def}) that Eq.\ (\ref{gamma-def}) obeys the Bose commutation relations
\begin{align}
[\hat{\tilde\gamma}_{{\bf k}},\hat{\tilde\gamma}_{{\bf k}'}^\dagger]= \delta_{{\bf k}{\bf k}'}, \hspace{5mm} [\hat{\tilde\gamma}_{{\bf k}},\hat{\tilde\gamma}_{{\bf k}'}]= 0.
\label{gamma-commutation}
\end{align}
The inverse of Eq.\ (\ref{gamma-def}) is easily obtained as
\begin{align}
\hat{\tilde c}_{{\bf k}} = u_{{\bf k}}\hat{\tilde\gamma}_{{\bf k}}+ v_{{\bf k}} \hat{\tilde\gamma}_{-{\bf k}}^{\dagger} .
\label{c-gamma}
\end{align}
The ket presented by Girardeau and Arnowitt\cite{GA59}
is given by a unitary transformation on $|{\cal N}\rangle_{\bf 0}$ that appears to be different from Eq.\ (\ref{|Phi_GA>}).
However, their equivalence can be confirmed by noting that both are (i) normalized and (ii) characterized as the vacuum of 
$\hat{\tilde\gamma}_{{\bf k}}$.

Evaluation of the ground-state energy using $|\Phi_{\rm GA}\rangle$ can be performed straightforwardly
as outlined in Appendix \ref{App:A_GA}.
Since the relevant expression is reproducible as a limit of the generalized version given below in Sect.\ \ref{subsec:3/2}, we do not carry it out here. 
It may suffice to point out here that 
\begin{align}
{\cal E}_{\rm GA}\equiv &\, \langle\Phi_{\rm GA}|\hat H|\Phi_{\rm GA}\rangle 
\notag \\
=&\,\langle\Phi_{\rm GA}|(\hat H_0+\hat H_1+\hat H_2)|\Phi_{\rm GA}\rangle ,
\label{E_GA}
\end{align}
i.e., $\langle\Phi_{\rm GA}|\hat H_{3/2}|\Phi_{\rm GA}\rangle=0$,
among the terms in Eq.\ (\ref{H-decomp}).
Neglecting the contribution of $\hat H_2$ in Eq.\ (\ref{E_GA}) corresponds to the Bogoliubov theory with a gapless excitation spectrum.\cite{Bogoliubov47,AGD63}
Inclusion of the $\hat H_2$ contribution, which is supposed to improve the variational wavefunction, nevertheless gives rises to an unphysical energy gap in the
excitation spectrum\cite{GA59,Griffin96} in contradiction to Goldstone's theorem I.\cite{GSW62,Weinberg96}

\subsection{Including $3/2$-body correlations\label{subsec:3/2}}

Now, we  improve $|\Phi_{\rm GA}\rangle$ so that $\hat H_{3/2}$ yields a finite contribution to lower the variational energy further below ${\cal E}_{\rm GA}$.
First, we introduce an operator $\hat{\tilde\pi}_3^\dagger$ given by
\begin{align}
\hat{\tilde\pi}_3^\dagger\equiv &\,\frac{1}{3!}
{\sum_{{\bf k}_1{\bf k}_2{\bf k}_3}}' w_{{\bf k}_1{\bf k}_2{\bf k}_3}\hat{\tilde\gamma}_{{\bf k}_1}^\dagger\hat{\tilde\gamma}_{{\bf k}_2}^\dagger\hat{\tilde\gamma}_{{\bf k}_3}^\dagger  ,
\label{pi_3}
\end{align}
where $w_{{\bf k}_1{\bf k}_2{\bf k}_3}$ is a variational parameter that is symmetric in $({\bf k}_1,{\bf k}_2,{\bf k}_3)$ by definition.
Using Eqs.\ (\ref{|Phi_GA>}) and (\ref{pi_3}), we construct the following wave function:
\begin{align}
|\Phi\rangle \equiv &\, {\cal A}_3 \exp\left(\hat{\tilde\pi}_3^\dagger\right) |\Phi_{\rm GA}\rangle ,
\label{|Phi>}
\end{align}
where ${\cal A}_3$ is determined by $\langle\Phi|\Phi\rangle=1$.

The variational  ground-state energy 
\begin{align}
{\cal E}\equiv \langle\Phi|\hat H|\Phi\rangle=\langle\Phi|(\hat H_0+\hat H_1+\hat H_{3/2}+\hat H_2)|\Phi\rangle
\label{calE-def}
\end{align}
can be estimated as follows.
First, we insert either $\hat\beta_{\bf 0}^\nu(\hat\beta_{\bf 0}^\dagger)^\nu=1$ or 
$(\hat\beta_{\bf 0}^\dagger)^\nu\hat\beta_{\bf 0}^\nu=1$ with $\nu= 1,2$ appropriately into Eqs.\ (\ref{H_1})-(\ref{H_2}) to express them in terms of 
$(\hat{\tilde c}_{{\bf k}}^\dagger,\hat{\tilde c}_{{\bf k}})$ as
\begin{subequations}
\label{H-decomp3}
\begin{align}
\hat H_{1}\equiv &\, {\sum_{{\bf k}}}'  \varepsilon_k\hat{\tilde c}_{{\bf k}}^\dagger\hat{\tilde c}_{{\bf k}}
+\frac{1}{{\cal V}}{\sum_{{\bf k}}}' \!\left(U_{0}+U_{k}\right)\hat c_{\bf 0}^\dagger\hat c_{\bf 0}\hat{\tilde c}_{{\bf k}}^\dagger\hat{\tilde c}_{{\bf k}}
\notag \\
&\,
+\frac{1}{2{\cal V}}{\sum_{{\bf k}}}'U_{k}\!\left(\hat c_{\bf 0}^\dagger\hat c_{\bf 0}^\dagger\hat\beta_{\bf 0}\hat\beta_{\bf 0}\hat{\tilde c}_{{\bf k}}\hat{\tilde c}_{-{\bf k}}
+\hat{\tilde c}_{-{\bf k}}^\dagger\hat{\tilde c}_{{\bf k}}^\dagger\hat\beta_{\bf 0}^\dagger\hat\beta_{\bf 0}^\dagger\hat c_{\bf 0}\hat c_{\bf 0} \!\right)\!,
\label{H_1-2}
\\
\hat H_{3/2}\equiv&\,\frac{1}{{\cal V}} {\sum_{{\bf k}_1{\bf k}_2{\bf k}_3}}' \delta_{{\bf k}_1+{\bf k}_2+{\bf k}_3,{\bf 0}}
U_{k_1}\!\left(\hat c_{\bf 0}^\dagger\hat\beta_{\bf 0}\hat{\tilde c}_{-{\bf k}_3}^\dagger\hat{\tilde c}_{{\bf k}_2}\hat{\tilde c}_{{\bf k}_1}
\right.
\notag \\
&\,\left.
+\hat{\tilde c}_{{\bf k}_1}^\dagger\hat{\tilde c}_{{\bf k}_2}^\dagger\hat{\tilde c}_{-{\bf k}_3}\hat\beta_{\bf 0}^\dagger\hat c_{\bf 0}\right)\! ,
\label{H_3/2-2}
\\
\hat H_{2}\equiv&\, \frac{1}{2{\cal V}}{\sum_{{\bf k}{\bf k}'{\bf q}}}'U_q
\hat{\tilde c}_{{\bf k}+{\bf q}}^\dagger\hat{\tilde c}_{{\bf k}'-{\bf q}}^\dagger\hat{\tilde c}_{{\bf k}'}\hat{\tilde c}_{{\bf k}} .
\label{H_2-2}
\end{align}
\end{subequations}
Then, we substitute the approximation
\begin{align}
(\hat{c}_{\bf 0}^\dagger)^n \hat{c}_{\bf 0}^m\approx  {\cal N}_{\bf 0}^{(n+m)/2}(\hat\beta_{\bf 0}^\dagger)^n \hat\beta_{\bf 0}^m
\label{c_0-approx}
\end{align}
in Eqs.\ (\ref{H_0}) and (\ref{H-decomp3}) with ${\cal N}_{\bf 0}$ denoting the number of condensed particles, 
 and use  $\hat\beta_{\bf 0}^\nu(\hat\beta_{\bf 0}^\dagger)^\nu=(\hat\beta_{\bf 0}^\dagger)^\nu\hat\beta_{\bf 0}^\nu=1$ to eliminate 
 $(\hat\beta_{\bf 0}^\dagger,\hat\beta_{\bf 0})$ from
the Hamiltonian in Eq.\ (\ref{calE-def}). 
The expectations of
the remaining $(\hat{\tilde c}_{{\bf k}}^\dagger,\hat{\tilde c}_{{\bf k}})$ operators can be calculated by performing the transformation of Eq.\ (\ref{c-gamma})
and using Eq.\ (\ref{gamma|Phi_GA>=0}), as detailed in Appendix \ref{App:A_3}.
Specifically, we obtain
\begin{subequations}
\label{nFW}
\begin{align}
\rho_{{\bf k}}\equiv&\,\langle\Phi|\hat{\tilde c}_{{\bf k}}^\dagger\hat{\tilde c}_{{\bf k}}|\Phi\rangle
\notag \\
=&\, |v_{{\bf k}}|^2\!\left(1+\frac{1}{2}{\sum_{{\bf k}_2{\bf k}_3}}'|w_{-{\bf k}{\bf k}_2{\bf k}_3}|^2\right)\!
+\frac{|u_{{\bf k}}|^2}{2}{\sum_{{\bf k}_2{\bf k}_3}}'|w_{{\bf k}{\bf k}_2{\bf k}_3}|^2 ,
\label{rho_k}
\end{align}
\begin{align}
F_{{\bf k}} \equiv&\, \langle\Phi|\hat{\tilde c}_{{\bf k}}\hat{\tilde c}_{-{\bf k}}|\Phi\rangle
\notag \\
=&\,u_{{\bf k}}v_{{\bf k}}\!\left(1+\frac{1}{2}{\sum_{{\bf k}_2{\bf k}_3}}'|w_{{\bf k}{\bf k}_2{\bf k}_3}|^2+\frac{1}{2}{\sum_{{\bf k}_2{\bf k}_3}}'|w_{-{\bf k}{\bf k}_2{\bf k}_3}|^2\right),
\label{F_k}
\end{align}
\begin{align}
W_{{\bf k}_1{\bf k}_2;{\bf k}_3}\equiv &\,\langle\Phi|\hat{\tilde c}_{-{\bf k}_3}^\dagger \hat{\tilde c}_{{\bf k}_2}\hat{\tilde c}_{{\bf k}_1}|\Phi\rangle
\notag \\
=&\, u_{{\bf k}_1}u_{{\bf k}_2}v_{{\bf k}_3}^*w_{{\bf k}_1{\bf k}_2{\bf k}_3}+v_{{\bf k}_1}v_{{\bf k}_2}u_{{\bf k}_3}w_{-{\bf k}_1-{\bf k}_2-{\bf k}_3}^* .
\label{barW}
\end{align}
\end{subequations}
The density of condensed particles ${\bar n}_{{\bf 0}}\equiv{\cal N}_{{\bf 0}}/{\cal V}$ is expressible using
the particle density ${\bar n}\equiv{\cal N}/{\cal V}$ and Eq.\ (\ref{rho_k}) as
\begin{align}
\bar n_{\bf 0}\equiv &\, \bar n-\frac{1}{{\cal V}}{\sum_{\bf k}}'\rho_{{\bf k}}.
\label{N_0}
\end{align}
Moreover, Eq.\ (\ref{calE-def}) is rewritten using Eqs.\ (\ref{nFW}) and (\ref{N_0}) as
\begin{align}
{\cal E}=&\, \frac{{\cal N}^2}{2{\cal V}} U_{0} +{\sum_{{\bf k}}}'\varepsilon_{{\bf k}}\rho_{{\bf k}}
+ \bar n_{\bf 0}{\sum_{{\bf k}}}' U_k\left(\rho_{{\bf k}}+\frac{F_{{\bf k}}+F_{{\bf k}}^*}{2}\right)
\notag \\
&\,+\frac{\sqrt{{\cal N}_{\bf 0}}}{{\cal V}}{\sum_{{\bf k}_1{\bf k}_2{\bf k}_3}}'\delta_{{\bf k}_1+{\bf k}_2+{\bf k}_3,{\bf 0}}
U_{k_1}\! \!\left( W_{{\bf k}_1{\bf k}_2;{\bf k}_3}
+W_{{\bf k}_1{\bf k}_2;{\bf k}_3}^*\right)
\notag \\
&\,
+\frac{1}{2{\cal V}}{\sum_{{\bf k}{\bf k}'}}'U_{|{\bf k}-{\bf k}'|}\left(\rho_{{\bf k}}\rho_{{\bf k}'}+F_{{\bf k}}F_{{\bf k}'}^*\right) ,
\label{calE}
\end{align}
where the first term results from collecting all the contributions proportional to $U_0$.
The Girardeau-Arnowitt functional of Eq.\ (\ref{E_GA}) is reproducible from Eq.\ (\ref{calE}) as
\begin{align}
{\cal E}_{\rm GA}={\cal E}[w_{{\bf k}_1{\bf k}_2{\bf k}_3}=0] .
\label{E_GA-calE}
\end{align}

\subsection{Stationarity conditions\label{subsec:sta-homo}}

To derive the stationarity conditions of Eq.\ (\ref{calE}), we assume the symmetries
\begin{align}
\phi_{{\bf k}}^*=\phi_{{\bf k}}, \hspace{5mm}
w_{{\bf k}_1{\bf k}_2{\bf k}_3}=w_{{\bf k}_1{\bf k}_2{\bf k}_3}^*=w_{-{\bf k}_1-{\bf k}_2-{\bf k}_3},
\label{phi-w-symm}
\end{align}
in the variational parameters.
Indeed, we will see that the symmetries are satisfied by the solutions.
The conditions $\delta{\cal E}/\delta\phi_{{\bf k}}=0$ and  $\delta{\cal E}/\delta w_{{\bf k}_1{\bf k}_2{\bf k}_3}=0$ for Eq.\ (\ref{calE}) 
can be calculated straightforwardly by performing the differentiations with the chain rule through the dependences in Eqs.\ (\ref{nFW}) and (\ref{N_0}),
where 
\begin{align}
{\sum_{{\bf k}_2{\bf k}_3}}'|w_{-{\bf k}{\bf k}_2{\bf k}_3}|={\sum_{{\bf k}_2{\bf k}_3}}'w_{{\bf k}{\bf k}_2{\bf k}_3}^2 
\notag
\end{align}
also holds owing to Eq.\ (\ref{phi-w-symm}).
We thereby find that $\delta{\cal E}/\delta\phi_{{\bf k}}=0$ and  $\delta{\cal E}/\delta w_{{\bf k}_1{\bf k}_2{\bf k}_3}=0$ yield
\begin{align}
2\xi_k\phi_{{\bf k}}+\Delta_k (\phi_{{\bf k}}^2+1)+\chi_k=0 ,
\label{phi_k-eq}
\end{align}
\begin{align}
w_{{\bf k}_1{\bf k}_2{\bf k}_3}=-\frac{b_{{\bf k}_1{\bf k}_2{\bf k}_3}}{a_{{\bf k}_1{\bf k}_2{\bf k}_3}},
\label{w_k-sol}
\end{align}
respectively.
The quantities $\xi_k\equiv \delta{\cal E}/\delta\rho_{{\bf k}}$, $\Delta_k\equiv \delta{\cal E}/\delta F_{{\bf k}}$,
and $\chi_k$ originating from the second line in Eq.\ (\ref{calE})
are given explicitly by
\begin{subequations}
\label{funcs}
\begin{align}
\xi_k\equiv &\, \varepsilon_k+\bar n_{{\bf 0}} U_k+ \frac{1}{{\cal V}}{\sum_{{\bf k}'}}'\bigl[(U_{|{\bf k}-{\bf k}'|}-U_{k'})\rho_{{\bf k}'}- U_{k'} F_{{\bf k}'}\bigr]
\notag \\
&\, -\frac{1}{{\cal V}\sqrt{{\cal N}_{{\bf 0}}}}{\sum_{{\bf k}_1{\bf k}_2{\bf k}_3}}'\delta_{{\bf k}_1+{\bf k}_2+{\bf k}_3,{\bf 0}}U_{k_1}W_{{\bf k}_1{\bf k}_2;{\bf k}_3},
\label{xi-def}
\end{align}
\begin{align}
\Delta_k\equiv  \bar n_{{\bf 0}} U_k+\frac{1}{{\cal V}}{\sum_{{\bf k}'}}'U_{|{\bf k}-{\bf k}'|} F_{{\bf k}'} ,
\label{Delta-def}
\end{align}
\begin{align}
\chi_k\equiv &\,\frac{2\sqrt{{\cal N}_{{\bf 0}}}}{\displaystyle1+ {\sum_{{\bf k}_2'{\bf k}_3'}}' |w_{{\bf k}{\bf k}_2'{\bf k}_3'}|^2}
\frac{1}{{\cal V}}{\sum_{{\bf k}_2{\bf k}_3}}'\delta_{{\bf k}+{\bf k}_2+{\bf k}_3,{\bf 0}}w_{{\bf k}{\bf k}_2{\bf k}_3}\frac{u_{{\bf k}_2}u_{{\bf k}_3}}{u_{{\bf k}}}
\notag \\
&\,
\times \! \bigl[U_{k_2}(1\!+\!\phi_{{\bf k}}\phi_{{\bf k}_2}\phi_{{\bf k}_3})+(U_{k}\!+\!U_{k_2})(\phi_{{\bf k}_2}\!+\!\phi_{{\bf k}}\phi_{{\bf k}_3})\bigr] ,
\label{chi-def}
\end{align}
and $a_{{\bf k}_1{\bf k}_2{\bf k}_3}$ and $b_{{\bf k}_1{\bf k}_2{\bf k}_3}$ in Eq.\ (\ref{w_k-sol}) denote
\begin{align}
a_{{\bf k}_1{\bf k}_2{\bf k}_3}\equiv&\, \sum_{j=1}^3 \left[\xi_{k_j}(2|v_{{\bf k}_j}|^2+1)+2\Delta_{k_j}u_{{\bf k}_j}v_{{\bf k}_j}\right],
\label{a-def}
\\
b_{{\bf k}_1{\bf k}_2{\bf k}_3}\equiv&\, \delta_{{\bf k}_1+{\bf k}_2+{\bf k}_3,{\bf 0}}\frac{\sqrt{{\cal N}_{\bf 0}}}{{\cal V}}u_{{\bf k}_1}u_{{\bf k}_2}u_{{\bf k}_3}
\notag \\
&\,\times
\bigl[ (U_{k_1}+U_{k_2})
(\phi_{{\bf k}_3}+\phi_{{\bf k}_1}\phi_{{\bf k}_2})
\notag \\
&\, +(U_{k_2}+U_{k_3})
(\phi_{{\bf k}_1}+\phi_{{\bf k}_2}\phi_{{\bf k}_3})
\notag \\
&\, 
+(U_{k_3}+U_{k_1})
(\phi_{{\bf k}_2}+\phi_{{\bf k}_3}\phi_{{\bf k}_1})\bigr] .
\label{b-def}
\end{align}
\end{subequations}
By imposing $\phi_{{\bf k}}\rightarrow 0$ for $k\rightarrow \infty$, Eq.\ (\ref{phi_k-eq}) can be transformed into
\begin{align}
\phi_{\bf k}=\frac{-\xi_k+\bigl[\xi_k^2-\Delta_k(\Delta_{k}+\chi_k)\bigr]^{1/2}}{\Delta_k} .
\label{phi_k-sol}
\end{align}
Equations (\ref{w_k-sol}) and (\ref{phi_k-sol}) with Eq.\ (\ref{funcs}) form a set of self-consistent equations that can be used 
to determine $\phi_{{\bf k}}$ and $w_{{\bf k}_1{\bf k}_2{\bf k}_3}$.
Equation (\ref{w_k-sol}) with Eqs.\ (\ref{a-def}) and (\ref{b-def}) indicates that $w_{{\bf k}_1{\bf k}_2{\bf k}_3}=O({\cal N}^{-1/2})$; thus, it is more convenient
 for numerical calculations to 
rewrite the whole expressions above in terms of $\tilde w_{{\bf k}_1{\bf k}_2{\bf k}_3}\equiv w_{{\bf k}_1{\bf k}_2{\bf k}_3}{\cal N}^{1/2}$.
This procedure also enables us to confirm that the terms with $w_{{\bf k}_1{\bf k}_2{\bf k}_3}$ in Eq.\ (\ref{calE}) make finite contributions in the thermodynamic limit.

\subsection{One-particle excitation spectrum}

Now, we study one-particle excitations from Eq.\ (\ref{|Phi>}) 
by calculating the first and second moments of the spectral function $A({\bf k},\varepsilon)$.\cite{TKK16}
As shown in Ref.\ \onlinecite{TKK16}, defining $A({\bf k},\varepsilon)$ using $(\hat{c}_{\bf k}^\dagger,\hat{c}_{\bf k})$
necessarily shifts the excitation spectrum by the chemical potential $\mu$, 
as expected naturally whenever adding a particle to the system.
To remove this undesirable shift, we here define the spectral function in terms of the number-conserving operators $(\hat{\tilde c}_{\bf k}^\dagger,\hat{\tilde c}_{\bf k})$
instead of  $(\hat{c}_{\bf k}^\dagger,\hat{c}_{\bf k})$.
The corresponding moments
\begin{align}
A_n({\bf k})\equiv \int_{-\infty}^\infty A({\bf k},\varepsilon)\varepsilon^n d\varepsilon
\end{align}
for $n=0,1,2$ can also be expressed as\cite{TKK16}
\begin{subequations}
\label{A_n}
\begin{align}
A_0({\bf k}) =&\,\langle\Phi|\hat{\tilde c}_{\bf k}\hat{\tilde c}_{\bf k}^\dagger|\Phi\rangle =1+\rho_{{\bf k}} ,
\label{A_0}
\\
A_1({\bf k}) =&\,\langle\Phi|[\hat{\tilde c}_{\bf k},\hat H]\hat{\tilde c}_{\bf k}^\dagger|\Phi\rangle,
\label{A_1}
\\
A_2({\bf k}) =&\,\langle\Phi|[\hat{\tilde c}_{\bf k},\hat H][\hat H,\hat{\tilde c}_{\bf k}^\dagger]|\Phi\rangle ,
\label{A_2}
\end{align}
\end{subequations}
where $\rho_{{\bf k}}$ is defined by Eq.\ (\ref{rho_k}).
The mean value and width of the one-particle excitation spectrum are obtained from the moments as\cite{TKK16}
\begin{subequations}
\label{barE's}
\begin{align}
\overline{E}_{{\bf k}}=&\,\frac{A_1({\bf k})}{A_0({\bf k})}, 
\label{barE's1}
\\
\overline{\Delta E}_{{\bf k}}=&\,\sqrt{\frac{A_2({\bf k})}{A_0({\bf k})}-\left[\frac{A_1({\bf k})}{A_0({\bf k})}\right]^2}.
\label{barE's2}
\end{align}
\end{subequations}
This $\overline{E}_{{\bf k}}$ in terms of $(\hat{\tilde c}_{\bf k}^\dagger,\hat{\tilde c}_{\bf k})$
represents the true excitation spectrum without the chemical-potential shift, unlike the definition with
$(\hat{c}_{\bf k}^\dagger,\hat{c}_{\bf k})$.
It is shown in Appendix \ref{App:A_12} that Eq.\ (\ref{barE's}) can be calculated straightforwardly but rather tediously. 
We thereby obtain the following expressions for the mean value and width of the one-particle spectrum:
\begin{subequations}
\label{barE-DE}
\begin{align}
\overline{E}_{{\bf k}}=&\,\xi_k+ \frac{\Delta_kF_{{\bf k}}}{1+\rho_{{\bf k}}}
\notag \\
&\,
+\frac{\sqrt{{\cal N}_{{\bf 0}}}}{(1+\rho_{{\bf k}}){\cal V}}{\sum_{{\bf k}_2{\bf k}_3}}'\delta_{{\bf k}+{\bf k}_2+{\bf k}_3,{\bf 0}}
w_{{\bf k}{\bf k}_2{\bf k}_3}u_{{\bf k}}u_{{\bf k}_2} u_{{\bf k}_3}
\notag \\
&\,\times  \bigl[U_{k_2}
(\phi_{{\bf k}}\!+\!\phi_{{\bf k}_2}\phi_{{\bf k}_3})
+(U_k\!+\!U_{k_2})(\phi_{{\bf k}_3}\!+\!\phi_{{\bf k}}\phi_{{\bf k}_2})\bigr] ,
\label{barE}
\\
\overline{\Delta E}_{{\bf k}}=&\,\Biggl\{\frac{(\bar nU_{k})^2}{2u_{\bf k}^4}{\sum_{{\bf k}_2{\bf k}_3}}' w_{{\bf k}{\bf k}_2{\bf k}_3}^2
+2 \bar n U_k\frac{\sqrt{{\cal N}}}{{\cal V}}{\sum_{{\bf k}_2{\bf k}_3}}' 
w_{{\bf k}{\bf k}_2{\bf k}_3}
\notag \\
&\,
\times\frac{u_{{\bf k}_2}u_{{\bf k}_3}}{u_{{\bf k}}^3} \left[U_{k_2} +(U_k+U_{k_2})\phi_{{\bf k}_2} \right]
\notag \\
&\,
+\frac{{\cal N}}{{\cal V}^2}{\sum_{{\bf k}_2{\bf k}_3}}' \delta_{{\bf k}+{\bf k}_2+{\bf k}_3,{\bf 0}}\frac{u_{{\bf k}_2}^2u_{{\bf k}_3}^2}{u_{\bf k}^2}
\left[ U_{k_2}(U_{k_2}+U_{k_3})
\right.
\notag \\
&\, 
+2(U_{k}+U_{k_2})(U_{k_2}+U_{k_3})\phi_{{\bf k}_2} 
+(U_{k}+U_{k_2})^2 \phi_{{\bf k}_2}^2
\notag \\
&\,\left. 
+(U_{k}+U_{k_2})(U_{k}+U_{k_3}) \phi_{{\bf k}_2} \phi_{{\bf k}_3} \right]\Biggr\}^{1/2} ,
\label{barDE}
\end{align}
\end{subequations}
where we set ${\cal N}_{{\bf 0}}\approx {\cal N}$ and  $A_0({\bf k})\approx u_{{\bf k}}^2$ in Eq.\ (\ref{barDE}) as justified
in the weak-coupling region noting Eqs.\ (\ref{rho_k}) and (\ref{N_0}).
Setting $w_{{\bf k}_1{\bf k}_2{\bf k}_3}=0$ in Eq.\ (\ref{barE}) reproduces 
the Girardeau-Arnowitt excitation spectrum,\cite{GA59} $E_{{\bf k}}^{\rm GA}=(\xi_k^2-\Delta_k^2)^{1/2}$, 
as confirmed by using Eq.\ (\ref{phi_k-sol}) with $\chi_k=0$ and  Eqs.\ (\ref{uv-def}) and (\ref{nFW}).
Moreover, the Bogoliubov spectrum\cite{Bogoliubov47} results from $E_{{\bf k}}^{\rm GA}$ by omitting the sums over ${\bf k}'$ in Eqs.\ (\ref{xi-def})
and (\ref{Delta-def}). Hence, our main interest in Eq.\ (\ref{barE}) is how the presence of $w_{{\bf k}_1{\bf k}_2{\bf k}_3}$ changes the one-particle spectrum. 
On the other hand, the last term on the right-hand side of Eq.\ (\ref{barDE}) indicates that 
incorporating the $3/2$-body processes gives rise to a finite width $\overline{\Delta E}_{{\bf k}}>0$ even for 
the excitations from the mean-field wave function with $w_{{\bf k}_1{\bf k}_2{\bf k}_3}=0$.

\subsection{Superposition over the number of condensed particles}

Finally, we derive an expression for the squared projection $\bigl|_0^{}\langle{\cal N}-n|\Phi\rangle\bigr|^2$ 
defined in terms of Eqs.\ (\ref{|n>_0-def}) and  (\ref{|Phi>}),
which enables us to study the superposition over the number of condensed particles in the wave function of Eq.\  (\ref{|Phi>}).
For this purpose, we note that $\hat{\tilde\pi}$ and $\hat{\tilde\pi}_3$ in Eqs.\ (\ref{|Phi_GA>}) and (\ref{|Phi>}) excite two and three particles from the condensate,
respectively.
With this observation, we expand the product ${\cal A}_{\rm GA}^{-2}{\cal A}_{3}^{-2}$ of Eqs.\ (\ref{A_GA^-2-1}) and (\ref{A_3^-2}) 
in a Taylor series and subsequently sort the terms according to the number of non-condensed particles.
Multiplying the resulting expression by
${\cal A}_{\rm GA}^2{\cal A}_{3}^2$, we obtain the squared projection for $n$ excitations as
\begin{align}
\left|^{}_0\hspace{-0.3mm}\langle{\cal N}-n|\Phi\rangle\right|^2=&\,
{\cal A}_{\rm GA}^2{\cal A}_{3}^2
\sum_{\{\ell_2,\ell_4,\cdots,\ell_3\}}
\delta_{n,2\ell_2+4\ell_4+\cdots+3\ell_3}
\notag \\
&\,\times\prod_{\lambda=1}^\infty \frac{{\cal I}_{2\lambda}^{\ell_{2\lambda}}}{\ell_{2\lambda}!}\frac{J_{3}^{\ell_3}}{\ell_3!},
\label{projection}
\end{align}
where the summation is performed over all the distinct sets of $\{\ell_1,\ell_2,\cdots,\ell_\nu\}$,
the quantities $I_{2\lambda}$ and $J_{3}$ are defined by
\begin{align}
I_{2\lambda}\equiv\frac{1}{2\lambda}{\sum_{{\bf k}}}'|\phi_{{\bf k}}|^{2\lambda} ,\hspace{5mm}
J_{3}\equiv\frac{1}{3!}{\sum_{{\bf k}_1{\bf k}_2{\bf k}_3}}'|w_{{\bf k}_1{\bf k}_2{\bf k}_3}|^2,
\label{IJ-def}
\end{align}
and we have omitted the contribution of $J_{3\lambda'}$ for $\lambda'\geq 2$ as being negligible in the weak-coupling region.
Equation (\ref{projection}) should obey the sum rule
\begin{align}
\sum_{n=0}^{\cal N}n \left|^{}_0\hspace{-0.3mm}\langle{\cal N}-n|\Phi\rangle\right|^2 ={\sum_{{\bf k}}}'\rho_{{\bf k}} ,
\label{sum-N_nc}
\end{align}
so as to be compatible with Eq.\ (\ref{N_0}). Equation (\ref{sum-N_nc}) can be used to check numerical results obtained with Eq.\ (\ref{projection}).

\section{Numerical Results\label{sec:NR}}

\subsection{Model potential and numerical procedures}

Numerical calculations were performed for the contact interaction potential $U_k=U$
used widely in the literature\cite{LHY57,BS57,AGD63}
to make a direct comparison possible.
For convenience, we express this $U$ alternatively as $4\pi\hbar^2 a_U/m$, i.e., 
\begin{align}
U_k=U=\frac{4\pi \hbar^2a_U}{m} .
\end{align}
The ultraviolet divergence inherent in the potential\cite{LHY57,BS57,AGD63} is removed by introducing a cutoff wavenumber $k_{\rm c}$
into every summation over ${\bf k}$ as 
\begin{align}
{\sum_{{\bf k}}}'\rightarrow{\sum_{{\bf k}}}'\theta(k_{\rm c}-k).
\end{align}
The $s$-wave scattering length $a$ of this interaction potential is obtained by\cite{LL-Q}
\begin{align}
\frac{m}{4\pi\hbar^2 a}=\frac{1}{U}+\int\frac{d^3k}{(2\pi)^3}\frac{\theta(k_{\rm c}-k)}{2\varepsilon_{k}} ,
\notag
\end{align}
which yields 
\begin{align}
a =\frac{a_U}{1+2k_{\rm c}a_U/\pi}.
\label{a-a_u}
\end{align}
We choose $k_{\rm c}$ so that $k_{\rm c}a_U\ll 1$ is satisfied, i.e., $a\approx a_U$ up to the leading order.

The characteristic energy and wavenumber of this system are given by 
\begin{align}
\varepsilon_U\equiv U\bar n  , \hspace{5mm} k_U\equiv \frac{\sqrt{2m\varepsilon_U}}{\hbar}=\sqrt{8\pi a_U \bar n},
\end{align}
respectively. They are used to transform
Eqs.\ (\ref{calE}) and (\ref{barE-DE}) into the dimensionless forms 
${\cal E}/{\cal N}\varepsilon_U$, $\bar E_{\bf k}/\varepsilon_U$, and $\overline{\Delta E}_{{\bf k}}/\varepsilon_U$ so that they are suitable for numerical calculations. 
Each sum over ${\bf k}$ in these quantities yields a factor of $(4/\pi)k_Ua_U=8(2a_U^3\bar n/\pi)^{1/2}$,
as seen by noting that $U/{\cal V}\varepsilon_U = 1/{\cal N}$ and 
\begin{align}
\frac{1}{{\cal N}}{\sum_{{\bf k}}}'=8\left(\frac{2a_U^3\bar n}{\pi}\right)^{1/2} \int_0^{\tilde k_{\rm c}}d\tilde k \, \tilde k^2 ,
\label{k-sum}
\end{align}
where $\tilde k\!\equiv\! k/k_U$.
Hence, Eq.\ (\ref{calE}) in the weak-coupling region is given as a series expansion in terms of $(a_U^3\bar n)^{1/2}\!\ll\! 1$.
In this context, its fourth term originating from $\hat H_{3/2}$ is of the same order as
the last one originating from $\hat H_2$ due to the presence of $\delta_{{\bf k}_1+{\bf k}_2+{\bf k}_3,{\bf 0}}$. 
Hence, process (c) of Fig.\ \ref{Fig1}  yields as important a contribution as the mean-field estimation of process (d),
meaning that it cannot be omitted even in the weak-coupling region.

Specifically, the energy per particle is expressible as
\begin{align}
\frac{{\cal E}}{{\cal N}}=&\,\frac{\varepsilon_U}{2}\biggl[1+\biggl(\frac{128}{15\sqrt{\pi}}-\frac{4\sqrt{2}}{\sqrt{\pi}}\tilde k_{{\rm c}}-\frac{4\sqrt{2}}{\sqrt{\pi} \tilde k_{\rm c}}\biggr)(a_U^3\bar n)^{1/2}
\notag \\
&\,
+2c_2 a_U^3\bar n\biggr]
\notag \\
\approx &\, \frac{2\pi\hbar^2 a \bar n}{m}\biggl[1+\biggl(\frac{128}{15\sqrt{\pi}}-\frac{4\sqrt{2}}{\sqrt{\pi} \tilde k_{\rm c}}\biggr)(a^3\bar n)^{1/2}+\cdots\biggr] .
\label{calE/calN}
\end{align}
The coefficient of $(a_U^3\bar n)^{1/2}$ in the first expression were obtained by (i) substituting the result of the Bogoliubov approximation
\begin{align}
\phi_{{\bf k}}^{\rm B}=-\frac{\varepsilon_k+\varepsilon_U-E_k^{\rm B}}{\varepsilon_U},\hspace{5mm}E_k^{\rm B}\equiv \sqrt{\varepsilon_k(\varepsilon_k+ 2\varepsilon_U)}
\label{phi_k^B}
\end{align}
for Eq.\ (\ref{phi_k-sol})
into the second and third terms of Eq.\ (\ref{calE}) with
$\bar n_{{\bf 0}}\rightarrow\bar n$, (ii) carrying out the integrations analytically, 
and (iii) performing an expansion in $1/{\tilde k}_{\rm c}$ up to the next-to-the-leading order. 
On the other hand, we used Eq.\ (\ref{a-a_u}) to derive the second expression for Eq.\ (\ref{calE/calN}) involving $a$, which for 
$\tilde k_{\rm c}\rightarrow\infty$ reduces  to the Lee-Huang-Yang expression 
for the ground-state energy.\cite{LHY57,BS57}
Since Eq.\ (\ref{calE/calN}) has a fairly strong $\tilde k_{\rm c}$ dependence, 
however, we use the first expression involving $a_U$ and
focus on the coefficient $c_2$  representing corrections beyond the Bogoliubov theory, 
to which $\hat H_{3/2}$ as well as $\hat H_{2}$ contributes.

The sums with $w_{{\bf k}_1{\bf k}_2{\bf k}_3}\!\propto\! \delta_{{\bf k}_1+{\bf k}_2+{\bf k}_3,{\bf 0}}$ in Eqs.\ (\ref{calE}) and (\ref{barE-DE}) 
were calculated by using the transformation
\begin{align}
&\,\frac{1}{{\cal N}}{\sum_{{\bf k}_2{\bf k}_3}}' \delta_{{\bf k}+{\bf k}_2+{\bf k}_3,{\bf 0}}
f(k,k_2,k_3)
\notag \\
=&\,8\left(\frac{2a_U^3\bar n}{\pi}\right)^{1/2}\frac{1}{2\tilde{k}}\int_0^{\tilde{k}_{\rm c}} d \tilde{k}_2 \tilde{k}_2
\int_{|\tilde{k}-\tilde{k}_2|}^{{\rm min}(\tilde{k}+\tilde{k}_2,\tilde k_{\rm c})}d \tilde{k}_{3}\tilde{k}_3
\notag \\
&\,\times  f(k,k_2,k_3) ,
\label{f-sum}
\end{align}
where we chose ${\bf k}$ along the $z$ direction,
expressed ${\bf k}_2=(k_2\sin\theta_2\cos\varphi_2,k_2\sin\theta_2\sin\varphi_2,k_2\cos\theta_2)$ in the polar coordinates, performed an integration over $0\!\leq\!\varphi_{2}\!\leq\! 2\pi$, 
and made a change of variables from $\theta_{2}$ to $k_3=(k^2+k_2^2+2kk_2\cos\theta_{2})^{1/2}$.
Integrals over $0\leq \tilde k\leq k_{\rm c}/k_U$ were calculated numerically  by making a change of variables $\tilde k=x^3$
to evaluate the important region $\tilde k\lesssim 1$ efficiently using a small number of integration points, $N_{\rm int}\lesssim 100$.

We solved Eqs.\  (\ref{w_k-sol}) and (\ref{phi_k-sol}) with Eq.\ (\ref{funcs}) iteratively from the initial values $\phi_{{\bf k}}=\phi_{{\bf k}}^{\rm B}$ and $w_{{\bf k}_1{\bf k}_2{\bf k}_3}=0$ to obtain $\phi_{{\bf k}}$ and $w_{{\bf k}_1{\bf k}_2{\bf k}_3}$ self-consistently, where
the condition $|\phi_{{\bf k}}|< 1$ in Eq.\ (\ref{uv-def}) was incorporated by expressing $\phi_{{\bf k}}=-\cos\theta_{{\bf k}}$.
The resulting solutions were substituted into Eqs.\ (\ref{calE}) and (\ref{barE-DE}) to calculate ${\cal E}/{\cal N}\varepsilon_U$, $\bar E_{\bf k}/\varepsilon_U$, 
and $\overline{\Delta E}_{{\bf k}}/\varepsilon_U$.

Equation (\ref{projection}), which represents the superposition over the number of condensed particles in our wave function $|\Phi\rangle$, 
was calculated by omitting the contribution of $\lambda>\lambda_{\rm c}$.
Choosing $\lambda_{\rm c}=10$ for ${\cal N}=20000$, $a_U^3 \bar n=1.0\times 10^{-6}$, and $k_{\rm c}/k_U=10$, we verified that the sum rule of Eq.\ (\ref{sum-N_nc}) 
was satisfied within $0.2\%$ by including $n\leq 70$.

\subsection{Results\label{subsec:NR}}

We now present our numerical results for $k_{\rm c}/k_U=10$, which corresponds to the cutoff energy $\varepsilon_{\rm c}=100\varepsilon_U$.
We varied the key parameter $a_U^3 \bar n$ between $10^{-10}$ and $10^{-5}$, where $k_{\rm c}a_U \ll 1$ is also satisfied
so that $a\approx a_U$ holds in Eq.\ (\ref{a-a_u}).

\begin{figure}[t]
        \begin{center}
                \includegraphics[width=0.9\linewidth]{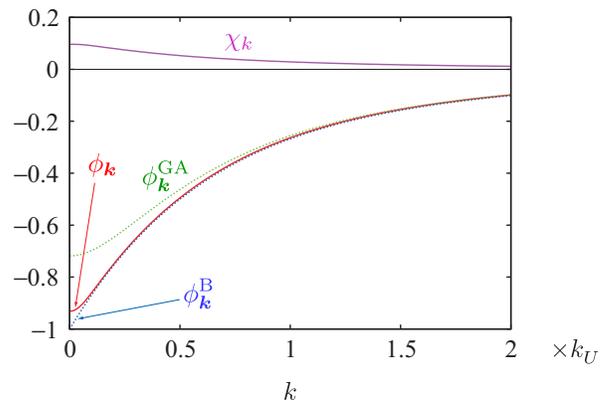}
        \end{center}
        \caption{ Plot of $\phi_{{\bf k}}$ and $\chi_k$ given by Eqs.\ (\ref{phi_k-sol}) and (\ref{chi-def}), respectively, as functions of $k$ for $a_U^3\bar n=1.0\times 10^{-6}$  and $k_{\rm c}/k_U=10$. 
        For comparison,  $\phi_{{\bf k}}^{\rm GA}$ and $\phi_{{\bf k}}^{\rm B}$ obtained by the Girardeau-Arnowitt and Bogoliubov theories, respectively, are also shown. \label{Fig2}}
\end{figure}

Figure \ref{Fig2} shows the $k$ dependence of  the basic functions $\phi_{{\bf k}}$ and $\chi_k$, Eqs.\ (\ref{phi_k-sol}) and (\ref{chi-def}), in comparison with 
 $\phi_{{\bf k}}^{\rm GA}$ and $\phi_{{\bf k}}^{\rm B}$ of the Girardeau-Arnowitt and Bogoliubov theories, respectively.
We observe that the $3/2$-body correlations bring the basic function $\phi_{{\bf k}}$ much closer to the prediction $\phi_{{\bf k}}^{\rm B}$ of 
the Bogoliubov theory than $\phi_{{\bf k}}^{\rm GA}$ by making $\chi_k$ finite.

\begin{figure}[t]
        \begin{center}
                \includegraphics[width=0.8\linewidth]{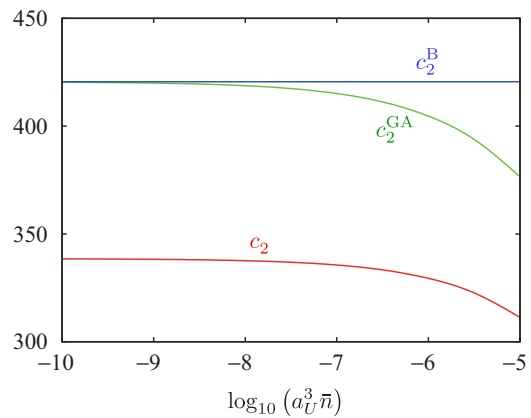}
        \end{center}
        \caption{Coefficient $c_2$ in Eq.\ (\ref{calE/calN}) in comparison with $c_2^{\rm GA}$ 
and $c_2^{\rm B}$ of the Girardeau-Arnowitt and Bogoliubov theories, respectively,
as functions of  $\log_{10}(a_U^3\bar n)$ for $k_{\rm c}/k_U=10$. \label{Fig3}}
\end{figure}

Figure \ref{Fig3} compares the coefficient $c_2$ in Eq.\ (\ref{calE/calN}) 
 with those of the Girardeau-Arnowitt theory ($\phi_{\bf k}\rightarrow\phi_{\bf k}^{\rm GA}$, $w_{{\bf k}_1{\bf k}_2{\bf k}_3}=0$) and the Bogoliubov theory ($\phi_{\bf k}\rightarrow\phi_{\bf k}^{\rm B}$, $w_{{\bf k}_1{\bf k}_2{\bf k}_3}=0$) as functions of $\log_{10}(a_U^3\bar n)$. 
It shows that the ground-state energy is $20\%$ less than the estimation by the  Girardeau-Arnowitt theory in the next-to-the-leading-order contribution.
Although the reduction is not large, this fact clearly indicates that the $3/2$-body correlations yield the same order of a contribution
as the mean-field interaction energy, meaning that they should be incorporated whenever the latter is included.
In other words, the mean-field approximation for BEC is quantitatively not effective even in the weak-coupling region.
Note that, in this context, the reduction of $c_2$ from $c_2^{\rm B}$ remains finite even for $a_U\rightarrow 0$, whereas the difference between
$c_2^{\rm GA}$ and $c_2^{\rm B}$ vanishes in the weak-coupling limit.

\begin{figure}[t]
        \begin{center}
                \includegraphics[width=0.9\linewidth]{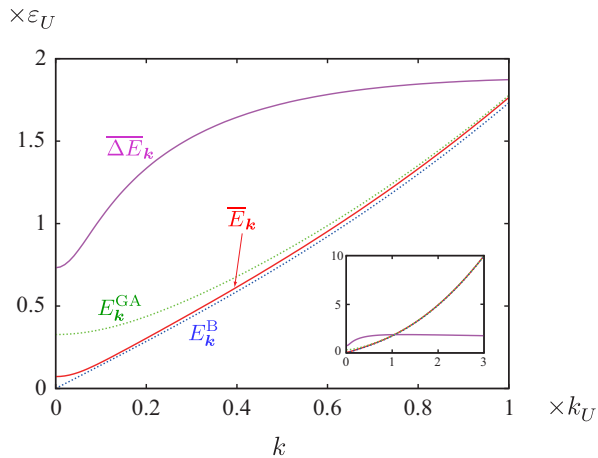}
        \end{center}
        \caption{ Plot of the mean value $\overline{E}_{{\bf k}}$ and width $\overline{\Delta E}_{{\bf k}}$ of the one-particle excitation spectrum given by Eqs.\ (\ref{barE}) and (\ref{barDE}), respectively, as functions of the wavenumber $k$ for $a_U^3\bar n=1.0\times 10^{-6}$  and $k_{\rm c}/k_U=10$. 
               The horizontal and vertical axes are normalized by $k_U$  and $\varepsilon_U$, respectively.
For comparison, the spectra   $E_{{\bf k}}^{\rm GA}$ and $E_{{\bf k}}^{\rm B}$ obtained by the Girardeau-Arnowitt  and  Bogoliubov
        theories,  respectively,  are also plotted.
         The inset shows the four curves over a wider range of $0\leq k\leq 3k_U$. \label{Fig4}}
\end{figure}

The $3/2$-body correlations also bring a {\em qualitative} change to the one-particle excitation spectrum from the mean-field description.
Figure \ref{Fig4} plots the mean value $\overline{E}_{\bf k}$ and the standard deviation $\overline{\Delta E}_{\bf k}$ of the one-particle excitation spectrum
for $a_U^3\bar n=1.0\times 10^{-6}$ and $k_{\rm c}/k_U=10$
calculated by Eqs.\ (\ref{barE}) and (\ref{barDE}), respectively.
The one-particle excitation has a finite lifetime $\tau_{\bf k}=\hbar/\overline{\Delta E}_{\bf k}<\infty$ even for $k\rightarrow 0$,
contrary to the predictions of the Girardeau-Arnowitt and Bogoliubov theories,\cite{Bogoliubov47,GA59}
due to the $3/2$-body processes of Fig.\ \ref{Fig1}(c).
These processes also have the effect of reducing the mean value $\overline{E}_{\bf k}$, which roughly represents the peak of the excitation spectrum, from the 
Girardeau-Arnowitt spectrum  $E_{{\bf k}}^{\rm GA}$ towards the Bogoliubov spectrum $E_{{\bf k}}^{\rm B}$.
The reduction becomes larger for $k\rightarrow 0$, but $\overline{E}_{\bf k}$ finally approaches a finite value because ${\Delta E}_{\bf k}>0$ even for $k\rightarrow0$.
The finite width ${\Delta E}_{\bf k}>0$ can be traced to the dynamical exchange of particles between the non-condensate reservoir and condensate;
its decrease for $k\rightarrow 0$ may be caused by the reduction of the available phase space.
The inset in Fig.\ \ref{Fig4} shows the four curves over a wider range of $0\leq k\leq 3k_U$. The peak $\overline{E}_{{\bf k}}$, as well as  $E_{{\bf k}}^{\rm GA}$ and 
$E_{{\bf k}}^{\rm B}$, approaches $\varepsilon_k$ for $k\gtrsim k_U$, but the width $\overline{\Delta E}_{\bf k}$ remains finite and decreases slowly
for $k\gtrsim k_U$. 

\begin{figure}[t]
        \begin{center}
                \includegraphics[width=0.9\linewidth]{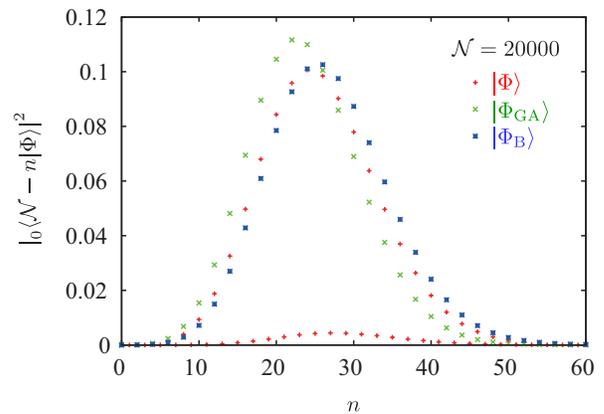}
        \end{center}
        \caption{ The squared projection $|_0\langle{\cal N}-n|\Phi\rangle|^2$ defined in terms of  Eqs.\ (\ref{|n>_0-def}) and (\ref{|Phi>}) as a function 
        of the number of non-condensed particles $n$ for ${\cal N}=20000$, $a_U^3\bar n=1.0\times 10^{-6}$,  and $k_{\rm c}/k_U=10$. 
        For comparison, the corresponding quantities obtained with the Bogoliubov and Girardeau-Arnowitt approximations
        are also plotted. \label{Fig5}}
\end{figure}

Finally, Fig.\ \ref{Fig5} shows the squared projection of Eq.\ (\ref{projection}) calculated for  ${\cal N}=20000$, $a_U^3\bar n=1.0\times 10^{-6}$,  and $k_{\rm c}/k_U=10$ 
by using Eq.\ (\ref{projection}).
The superposition over the number of condensed particles has a peak at $n=24$ and becomes negligible for $n\gtrsim 60$.
The profile with even integers is close to the one by the Bogoliubov approximation.
On the other hand, there also is an extra contribution from odd integers due to
the $3/2$-body correlations.
It should be emphasized that the superposition is here realized {\it physically and naturally} due to the interaction,
contrary to the case of photons 
with no interactions,  for which 
Sudarshan\cite{Sudarshan63} and Glauber\cite{Glauber63} introduced the superposition {\it mathematically} to describe their coherence.
Thus,  the interaction plays a crucial role in establishing the superposition indispensable for the phase coherence
within fixed-number Bose-Einstein condensates, which is also maintained here {\it dynamically} by the $3/2$-body processes.

\section{Summary\label{sec:summary}}

We have constructed a variational wave function for the ground state of weakly interacting bosons given by Eq.\ (\ref{|Phi>}).
It incorporates the $3/2$-body processes of Fig.\ \ref{Fig1}(c) to give a lower energy than the mean-field Girardeau-Arnowitt wave function,
as shown in Fig.\ \ref{Fig3}.
This wave function is given as a superposition in terms of the number of condensed particles, as seen in Fig.\ \ref{Fig5}, 
where non-condensed particles serve as a particle reservoir in the fixed-number formalism.
Thus, the interaction naturally brings a superposition indispensable for coherence\cite{Anderson66} to the condensate, 
which is sustained here temporarily by the dynamical $3/2$-body processes.
The corresponding excitation spectrum is characterized by a finite lifetime even in the long-wavelength limit,
as seen in Fig.\ \ref{Fig4}, which reflects the dynamical exchange of particles between the non-condensate reservoir and condensate
by the $3/2$-body processes. The unphysical energy gap appearing in the mean-field treatment\cite{GA59} seems removed
by the resulting broadening of the spectrum to give a finite spectral weight around $\varepsilon=0$ for $k\rightarrow 0$.
However, it is still not clear whether or not Goldstone's theorem I, which is given in terms of Green's function for the non-condensate, 
is satisfied by the present treatment.
This issue remains to be clarified in the future by developing a formalism to describe the one-particle excitations 
with the $3/2$-body correlations in terms of Green's function.

It is widely accepted that the equilibrium in thermodynamics is realized and sustained by the exchange of momenta through collisions of particles.
In contrast, little attention seems to have been paid to the origin of the coherence in Bose-Einstein condensates and superconductors.
The present paper has clarified the key role played by the interaction in realizing the superposition over the number of condensed particles
 indispensable for coherence. An observation of the finite lifetime in the one-particle excitation spectrum,
which has been predicted in the previous\cite{TKK16} and present papers,
will provide a definite confirmation that the coherence is maintained dynamically.

\appendix

\section{Calculations of ${\cal A}_{\rm GA}^{-2}$ and ${\cal E}_{\rm GA}$\label{App:A_GA}}

\begin{figure}[b]
\begin{center}
\includegraphics[width=0.95\linewidth]{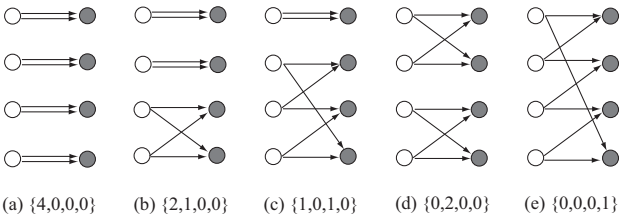}
\caption{Diagrammatic representations of $Q_4$ with five distinct sets of $\{\ell_1,\ell_2,\ell_3,\ell_4\}$
in the summation of Eq.\ (\ref{Q_nu}). 
An open (filled) circle with two outgoing (incoming) arrows denotes $\hat\pi^\dagger$ ($\hat\pi$).
\label{FigA1}}
\end{center}
\end{figure} 

The normalization constant ${\cal A}_{\rm GA}$ in Eq.\ (\ref{|Phi_GA>}) plays a key role in the evaluation of Eq.\ (\ref{E_GA}). 
Hence, we start by deriving its analytic expression.
Imposing the condition $\langle\Phi_{\rm GA}|\Phi_{\rm GA}\rangle=1$ on Eq.\ (\ref{|Phi_GA>}) yields 
\begin{subequations}
\label{A_GA-cal}
\begin{align}
{\cal A}_{\rm GA}^{-2}=&\, \sum_{\nu=0}^{[{\cal N}/2]} Q_\nu,\hspace{5mm} Q_\nu \equiv\frac{\langle 0 | \hat\pi^\nu (\hat\pi^\dagger)^{\nu}|0\rangle}{(\nu!)^2} .
\label{A_GA^-2-def}
\end{align}
The quantity $\langle 0 | \hat\pi^\nu (\hat\pi^\dagger)^{\nu}|0\rangle$ can be evaluated by using\cite{Kita-Text}
\begin{align}
\langle 0|\hat c_{{\bf k}_1'}\cdots \hat c_{{\bf k}_{2\nu}'}\hat c_{{\bf k}_{2\nu}}^\dagger\cdots\hat c_{{\bf k}_1}^\dagger|0\rangle=\sum_{\hat P}\prod_{j=1}^{2\nu}\delta_{{\bf k}_{p_j}'{\bf k}_{j}},
\label{Wick}
\end{align}
where $\hat P$ is a permutation $j\!\rightarrow \!p_j$ with $2\nu$ elements.\cite{Kita-Text}
Various terms in $\langle 0 | \hat\pi^\nu (\hat\pi^\dagger)^{\nu}|0\rangle$ can be classified diagrammatically 
according to the number of connected subgroups, as exemplified for $\nu=4$ in Fig.\ \ref{FigA1}.
Using the diagrams, we obtain an analytic expression for $Q_\nu$ as
\begin{align}
Q_\nu 
=&\, 
\sum_{\{\ell_1,\ell_2,\cdots,\ell_\nu\}}\frac{\delta_{\ell_1+2\ell_2+\cdots+ \nu\ell_\nu,\nu}}{(\nu!)^2}
\notag \\
&\,\times
 \left[\frac{\nu!}{\ell_1!(1!)^{\ell_1}\ell_2!(2!)^{\ell_2}\cdots\ell_\nu!(\nu!)^{\ell_\nu}}\right]^2
\notag \\
&\,\times  \ell_1!\ell_2!\cdots\ell_\nu! \prod_{\lambda=1}^\nu \left[\langle 0 | \hat\pi^\lambda (\hat\pi^\dagger)^{\lambda}|0\rangle_{\rm c}\right]^{\ell_\lambda} ,
\label{Q_nu}
\end{align}
where the summation is performed over all the distinct sets of $\{\ell_1,\ell_2,\cdots,\ell_\nu\}$.
Specifically, the factor in the large square brackets of Eq.\ (\ref{Q_nu}) denotes the number of combinations for distributing $\nu$ persons (i.e., $\hat\pi$ or $\hat\pi^\dagger$) into 
$(\ell_1,\ell_2,\cdots,\ell_\nu)$ rooms, where $\ell_\lambda$ ($\lambda=1,2,\cdots,\nu$) is the number of rooms with $\lambda$ beds.
Factor $\ell_\lambda!$ after the square brackets is the number of combinations in forming $\ell_\lambda$ pairs of 
$\bigl[\bigl(\hat\pi^\dagger\bigr)^\lambda,\hat \pi^\lambda\bigr]$ to construct a {\it connected} expectation 
$\langle 0 | \hat\pi^\lambda (\hat\pi^\dagger)^{\lambda}|0\rangle_{\rm c}$ for each pair.
Substituting Eq.\ (\ref{Q_nu}) into Eq.\ (\ref{A_GA^-2-def}), we obtain
\begin{align}
{\cal A}_{\rm GA}^{-2}=&\, \sum_{\nu=0}^{[{\cal N}/2]}
\sum_{\{\ell_1,\ell_2,\cdots,\ell_\nu\}}\delta_{\ell_1+2\ell_2+\cdots+ \nu\ell_\nu,\nu}
\notag \\
&\,\times
\prod_{\lambda=1}^\nu \frac{1}{\ell_\lambda!}\left[\frac{\langle 0 | \hat\pi^\lambda (\hat\pi^\dagger)^{\lambda}|0\rangle_{\rm c}}{(\lambda!)^{2}}\right]^{\ell_\lambda} 
\notag \\
\approx &\,
\exp\left( \sum_{\lambda=1}^\infty I_{2\lambda} \right) ,\hspace{5mm}I_{2\lambda}\equiv  \frac{\langle 0 | \hat\pi^\lambda (\hat\pi^\dagger)^{\lambda}|0\rangle_{\rm c}}{(\lambda!)^{2}},
\label{A_GA^-2-1}
\end{align}
\end{subequations}
where we have replaced the upper limit $[{\cal N}/2]$ by $\infty$ to derive the second expression based on the observation that
$Q_\nu$ for $\nu\!\sim \!{\cal N}/2$ can be set equal to zero in the weak-coupling region; see Fig.\ \ref{Fig5} regarding this point. 
The connected expectations $(I_{2},I_4,I_6,\cdots)$ 
have the common diagrammatic structure shown in Fig.\ \ref{FigA1}(e) for $\lambda=4$ and are expressible generally as
\begin{align}
I_{2\lambda}=\frac{2^{2\lambda-1}\lambda!(\lambda-1)!}{2^{2\lambda}(\lambda!)^2}{\sum_{\bf k}}'|\phi_{\bf k}|^{2\lambda}
=\frac{1}{2\lambda}{\sum_{\bf k}}'|\phi_{\bf k}|^{2\lambda} .
\label{<0|pipi|0>_c}
\end{align}
Here the factor $2^{2\lambda-1}\lambda!(\lambda-1)!$ originates from the number of combinations in connecting the $2\lambda$ pairs of field operators.
Substituting Eq.\ (\ref{<0|pipi|0>_c}) into Eq.\ (\ref{A_GA^-2-1}) yields
\begin{align}
{\cal A}_{\rm GA}^{-2}=&\,\exp\left[ -\frac{1}{2}{\sum_{\bf k}}'\ln (1-|\phi_{\bf k}|^{2})\right] 
\notag \\
=&\,\exp\left( {\sum_{\bf k}}' \ln\, u_{{\bf k}}\,\right) ,
\label{A_GA^-2}
\end{align}
where we used Eq.\ (\ref{uv-def}).

This quantity ${\cal A}_{\rm GA}^{-2}$ enables us to calculate various expectations with Eq.\ (\ref{|Phi_GA>}).
First, $\langle\Phi_{\rm GA}|\hat c_{{\bf k}}^\dagger\hat c_{{\bf k}}|\Phi_{\rm GA}\rangle$ for ${\bf k}\neq 0$ can be transformed as
\begin{align}
 \langle\Phi_{\rm GA}|\hat c_{{\bf k}}^\dagger\hat c_{{\bf k}}|\Phi_{\rm GA}\rangle 
=&\, {\cal A}_{\rm GA}^{2}\sum_{\nu=0}^{[{\cal N}/2]} 
\frac{\langle 0 |\hat \pi^\nu \hat c_{{\bf k}}^\dagger\bigl[\hat c_{{\bf k}},(\hat\pi^\dagger)^\nu\bigr]|0\rangle}{(\nu!)^2}
\notag \\
=&\,  {\cal A}_{\rm GA}^{2}\sum_{\nu=0}^{[{\cal N}/2]}\frac{\langle 0 |\hat \pi^\nu \hat c_{{\bf k}}^\dagger\hat c_{-{\bf k}}^\dagger(\hat\pi^\dagger)^{\nu-1}|0\rangle}{(\nu!)^2}\nu \phi_{{\bf k}}
\notag \\
=&\,{\cal A}_{\rm GA}^{2}\sum_{\nu=0}^{[{\cal N}/2]}\frac{\delta Q_\nu}{\delta \phi_{{\bf k}} }\phi_{{\bf k}}
= \phi_{{\bf k}}\frac{\delta \ln {\cal A}_{\rm GA}^{-2}}{\delta \phi_{{\bf k}}}
\notag \\
 \approx &\,  \phi_{{\bf k}}\frac{\phi_{{\bf k}}^*}{1-|\phi_{{\bf k}}|^2}
=|v_{{\bf k}}|^2 .
\label{n_qq'}
\end{align}
Here we used $\hat{c}_{{\bf k}}|0\rangle=0$ and Eq.\ (\ref{[tc,tpi]}) for the first two equality signs, 
then expressed $\langle 0 |\hat \pi^\nu \hat c_{{\bf k}}^\dagger\hat c_{-{\bf k}}^\dagger(\hat\pi^\dagger)^{\nu-1}|0\rangle\nu$ 
in terms of a functional derivative of $Q_\nu$ in Eq.\ (\ref{A_GA^-2-def}) noting Eq.\ (\ref{pi-def}) and that $\phi_{-{\bf k}}=\phi_{{\bf k}}$,
and finally used Eqs.\ (\ref{A_GA^-2}) and (\ref{uv-def}) for the last two equality signs. 
Note that Eq.\ (\ref{n_qq'}) can be derived more easily by expressing $\hat c_{{\bf k}}^\dagger\hat c_{{\bf k}}=\hat{\tilde c}_{{\bf k}}^\dagger\hat{\tilde c}_{{\bf k}}$, 
performing the transformation of Eq.\ (\ref{c-gamma}), and using Eqs.\ (\ref{gamma|Phi_GA>=0}) and (\ref{gamma-commutation}).
Second, the number of condensed particles can be estimated as
\begin{align}
{\cal N}_{\bf 0}\equiv&\, \langle\Phi_{\rm GA}|\hat c_{\bf 0}^\dagger\hat c_{\bf 0}|\Phi_{\rm GA}\rangle={\cal A}_{\rm GA}^{2}\sum_{\nu=0}^{[{\cal N}/2]} ({\cal N}-2\nu) Q_\nu
\notag \\
=&\,{\cal N}-2{\cal A}_{\rm GA}^{2}\sum_{\nu=0}^{[{\cal N}/2]}{\sum_{{\bf k}}}'\frac{1}{2}\phi_{{\bf k}}\frac{\delta Q_\nu}{\delta \phi_{{\bf k}}}
\notag \\
=&\, {\cal N}-{\sum_{{\bf k}}}'\phi_{{\bf k}}\frac{\delta \ln {\cal A}_{\rm GA}^{-2}}{\delta \phi_{{\bf k}}}
\notag \\
=&\,{\cal N}-{\sum_{{\bf k}}}' |v_{{\bf k}}|^2 .
\label{N_0-GA}
\end{align}
The second term in the final expression denotes the number of depleted particles, which can also be obtained from Eq.\ (\ref{n_qq'}) by summing it over ${\bf k}$. 
Third, the expectation of $\hat c_{\bf 0}^\dagger\hat c_{\bf 0}^\dagger\hat c_{\bf 0}\hat c_{\bf 0}\approx(\hat c_{\bf 0}^\dagger\hat c_{\bf 0})^2$ is calculated as
\begin{align}
&\, \langle\Phi_{\rm GA}|\hat c_{\bf 0}^\dagger\hat c_{\bf 0}^\dagger\hat c_{\bf 0}\hat c_{\bf 0}|\Phi_{\rm GA}\rangle
\approx {\cal A}_{\rm GA}^{2}\sum_{\nu=0}^{[{\cal N}/2]} ({\cal N}-2\nu)^2Q_\nu
\notag \\
=&\,{\cal N}^2\!-2^2{\cal N}{\sum_{{\bf k}}}'\frac{1}{2}\phi_{{\bf k}}\frac{\delta \ln {\cal A}_{\rm GA}^{-2}}{\delta \phi_{{\bf k}}}
\notag \\
&\,+2^2\left(\frac{1}{2}{\sum_{{\bf k}}}'\phi_{{\bf k}}\frac{\delta }{\delta \phi_{{\bf k}}}\right)^2\ln {\cal A}_{\rm GA}^{-2}
\notag \\
=&\, {\cal N}_{\bf 0}^2 + {\sum_{{\bf k}}}' \frac{|\phi_{{\bf k}}|^2}{(1-|\phi_{{\bf k}}|^2)^2} \approx  {\cal N}_{\bf 0}^2 ,
\label{N_0^2-GA}
\end{align}
where ``$\approx$'' implies neglecting terms of $O({\cal N})$ compared with those of $O({\cal N}^2)$.
Equations (\ref{N_0-GA}) and (\ref{N_0^2-GA}) justify the procedure of replacing $\hat c_{\bf 0}$ by $\sqrt{{\cal N}_{\bf 0}}$
in the variational calculation using the Girardeau-Arnowitt wave function.

\section{Calculations of ${\cal A}_3^{-2}$ and ${\cal E}$\label{App:A_3}}

\begin{figure}[b]
\begin{center}
\includegraphics[width=0.95\linewidth]{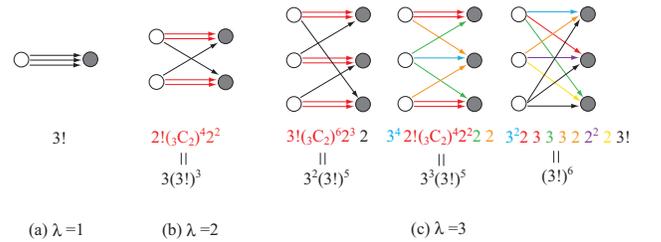}
\caption{ Diagrammatic representations of $J_{3\lambda}$ for $\lambda=1,2,3$. 
An open (filled) circle with three outgoing (incoming) arrows denotes $\hat{\tilde\pi}_3^\dagger$ ($\hat{\tilde\pi}_3$).
The weight below each figure denotes the number of combinations for realizing the connection.
\label{FigB1}}
\end{center}
\end{figure}

The transformation of ${\cal A}_{\rm GA}^{-2}$ in Eq.\ (\ref{A_GA-cal}) is also applicable to that of ${\cal A}_3^{-2}$ from $\langle\Phi|\Phi\rangle=1$.
Specifically, we only need to replace 
$(|0\rangle,\hat\pi,[{\cal N}/2])$ in Eq.\ (\ref{A_GA-cal}) by $(|\Phi_{\rm GA}\rangle,\hat{\tilde\pi}_3,\infty)$.
We thereby obtain
\begin{align}
\ln {\cal A}_3^{-2}= \sum_{\lambda=1}^\infty J_{3\lambda},\hspace{5mm}
J_{3\lambda}\equiv \frac{\langle\Phi_{\rm GA}| \hat{\tilde\pi}_3^\lambda (\hat{\tilde\pi}_3^\dagger)^{\lambda}|\Phi_{\rm GA}\rangle_{\rm c}}{(\lambda!)^{2}} .
\label{A_3^-2}
\end{align}
This quantity is analogous to Eq.\ (\ref{A_GA^-2-1}) with the correspondence $(\hat{\tilde\gamma}_{{\bf k}},|\Phi_{\rm GA}\rangle,3\lambda)\leftrightarrow (\hat c_{{\bf k}},|0\rangle,2\lambda)$.
Hence, we can also evaluate it analytically using Eq.\ (\ref{Wick}), the results of which can be classified diagrammatically as Fig.\ \ref{FigB1}.
In particular, the lowest-order contribution is obtained as
\begin{align}
J_3 =\frac{1}{3!}{\sum_{{\bf k}_1{\bf k}_2{\bf k}_3}}'|w_{{\bf k}_1{\bf k}_2{\bf k}_3}|^2   .
\label{calJ_3}
\end{align}
It turns out that the terms of $\lambda\!\geq\! 2$ in Eq.\ (\ref{A_3^-2}), which have increasing numbers of summations over ${\bf k}\!\neq \!{\bf 0}$, 
are negligible compared with Eq.\ (\ref{calJ_3}) in the weak-coupling region.
 
Equation (\ref{A_3^-2}) enables us to calculate various expectations in terms of $|\Phi\rangle$ in Eq.\ (\ref{|Phi>}).
Among them, the expectations of $\hat{\tilde c}_{{\bf k}}^\dagger\hat{\tilde c}_{{\bf k}}$ and $\hat{\tilde c}_{{\bf k}}\hat{\tilde c}_{-{\bf k}}$ for ${\bf k}\!\neq\! {\bf 0}$ are transformed 
by substituting Eq.\ (\ref{c-gamma}), using Eq.\ (\ref{gamma-commutation}) to arrange the quasiparticle operators into the normal order,
and noting that $\langle\Phi|\hat{\tilde\gamma}_{{\bf k}}\hat{\tilde\gamma}_{-{\bf k}}|\Phi\rangle=0$.
We thereby obtain expressions for $\rho_{{\bf k}}\!\equiv\! \langle\Phi|\hat{\tilde c}_{{\bf k}}^\dagger\hat{\tilde c}_{{\bf k}}|\Phi\rangle$ and
$F_{{\bf k}}\! \equiv \!\langle\Phi|\hat{\tilde c}_{{\bf k}}\hat{\tilde c}_{-{\bf k}}|\Phi\rangle$ as
\begin{subequations}
\label{cc-expectations}
\begin{align}
\rho_{{\bf k}}=&\ |v_{{\bf k}}|^2\left(1
+\langle\Phi|\hat{\tilde\gamma}_{-{\bf k}}^\dagger\hat{\tilde\gamma}_{-{\bf k}}|\Phi\rangle\right)
+u_{{\bf k}}^2\langle\Phi|\hat{\tilde\gamma}_{{\bf k}}^\dagger\hat{\tilde\gamma}_{{\bf k}}|\Phi\rangle ,
\label{c^dagger-c}
\\
F_{{\bf k}}=&\, u_{{\bf k}}v_{{\bf k}}\left(1 +\langle\Phi|\hat{\tilde\gamma}_{{\bf k}}^\dagger\hat{\tilde\gamma}_{{\bf k}}|\Phi\rangle
+\langle\Phi|\hat{\tilde\gamma}_{-{\bf k}}^\dagger\hat{\tilde\gamma}_{-{\bf k}}|\Phi\rangle\right) .
\end{align}
\end{subequations}
The expectation $\langle\Phi|\hat{\tilde\gamma}_{{\bf k}}^\dagger\hat{\tilde\gamma}_{{\bf k}}|\Phi\rangle $ in Eq.\ (\ref{cc-expectations}) can be transformed by using Eqs.\ 
(\ref{gamma|Phi_GA>=0}), (\ref{gamma-commutation}), (\ref{pi_3}), (\ref{|Phi>}), (\ref{A_3^-2}), and (\ref{calJ_3}) as
\begin{align}
&\,
\langle\Phi|\hat{\tilde\gamma}_{{\bf k}}^\dagger\hat{\tilde\gamma}_{{\bf k}}|\Phi\rangle
\notag \\
=&\,{\cal A}_3^2 \langle\Phi_{\rm GA}|\exp\left(\hat{\tilde\pi}_3\right)\hat{\tilde\gamma}_{{\bf k}}^\dagger\hat{\tilde\gamma}_{{\bf k}}\exp\left(\hat{\tilde\pi}_3^\dagger\right)|\Phi_{\rm GA}\rangle
\notag \\
=&\,{\cal A}_3^2 \langle\Phi_{\rm GA}|\exp\left(\hat{\tilde\pi}_3\right)\hat{\tilde\gamma}_{{\bf k}}^\dagger\left[\hat{\tilde\gamma}_{{\bf k}},\exp\left(\hat{\tilde\pi}_3^\dagger\right)\right]|\Phi_{\rm GA}\rangle
\notag \\
=&\,{\cal A}_3^2 \langle\Phi_{\rm GA}|\exp\left(\hat{\tilde\pi}_3\right)\hat{\tilde\gamma}_{{\bf k}}^\dagger\left[\hat{\tilde\gamma}_{{\bf k}},\hat{\tilde\pi}_3^\dagger\right]\exp\left(\hat{\tilde\pi}_3^\dagger\right)|\Phi_{\rm GA}\rangle
\notag \\
=&\,  {\sum_{{\bf k}_2{\bf k}_3}}'\frac{3w_{{\bf k}{\bf k}_2{\bf k}_3}}{3!}
{\cal A}_3^2 \langle\Phi|\hat{\tilde\gamma}_{{\bf k}}^\dagger\hat{\tilde\gamma}_{{\bf k}_2}^\dagger\hat{\tilde\gamma}_{{\bf k}_3}^\dagger|\Phi\rangle
\notag \\
=&\,{\sum_{{\bf k}_2{\bf k}_3}}' \frac{w_{{\bf k}{\bf k}_2{\bf k}_3}}{2}
{\cal A}_3^2\frac{\delta {\cal A}_3^{-2}}{\delta w_{{\bf k}{\bf k}_2{\bf k}_3}}
\notag \\
=&\,{\sum_{{\bf k}_2{\bf k}_3}}' \frac{w_{{\bf k}{\bf k}_2{\bf k}_3}}{2}
\frac{\delta\ln {\cal A}_3^{-2}}{\delta w_{{\bf k}{\bf k}_2{\bf k}_3}}
\approx  \frac{1}{2}{\sum_{{\bf k}_2{\bf k}_3}}'|w_{{\bf k}{\bf k}_2{\bf k}_3}|^2 .
\label{expectation1}
\end{align}
Substituting Eq.\ (\ref{expectation1}) into Eq.\ (\ref{cc-expectations}), we obtain Eqs.\ (\ref{rho_k}) and (\ref{F_k}).

We can also transform $W_{{\bf k}_1{\bf k}_2;{\bf k}_3}\!\equiv\!\langle\Phi|\hat{\tilde c}_{-{\bf k}_3}^\dagger \hat{\tilde c}_{{\bf k}_2}\hat{\tilde c}_{{\bf k}_1}|\Phi\rangle$ for ${\bf k}_1,{\bf k}_2,{\bf k}_3\!\neq\! 0$ 
by substituting Eq.\ (\ref{c-gamma}), using Eq.\ (\ref{gamma-commutation}) to arrange the quasiparticle operators into the normal order,
and noting that $\langle\Phi|\hat{\tilde\gamma}_{{\bf k}_1}^\dagger\hat{\tilde\gamma}_{{\bf k}_2}\hat{\tilde\gamma}_{{\bf k}_3}|\Phi\rangle=0$ into
\begin{align}
W_{{\bf k}_1{\bf k}_2;{\bf k}_3}
= &\,u_{{\bf k}_1}u_{{\bf k}_2}v_{{\bf k}_3}^*\langle\Phi|\hat{\tilde\gamma}_{{\bf k}_3}\hat{\tilde\gamma}_{{\bf k}_2}\hat{\tilde\gamma}_{{\bf k}_1}|\Phi\rangle
\notag \\
&\,
+v_{{\bf k}_1}v_{{\bf k}_2}u_{{\bf k}_3}\langle\Phi|
\hat{\tilde\gamma}_{-{\bf k}_3}^\dagger\hat{\tilde\gamma}_{-{\bf k}_2}^\dagger\hat{\tilde\gamma}_{-{\bf k}_1}^\dagger|\Phi\rangle.
\label{expectation3}
\end{align}
Now, the last three lines of Eq.\ (\ref{expectation1}) indicate $\langle\Phi|\hat{\tilde\gamma}_{{\bf k}_3}\hat{\tilde\gamma}_{{\bf k}_2}\hat{\tilde\gamma}_{{\bf k}_1}|\Phi\rangle
= w_{{\bf k}_1{\bf k}_2{\bf k}_3}$. Substituting this into Eq.\ (\ref{expectation3}), we obtain Eq.\ (\ref{barW}).

Finally, the expectation of the operator product in Eq.\ (\ref{H_2-2}), which has the highest order among the terms on the right-hand side of Eq.\ (\ref{H-decomp}),
can be evaluated most easily by the Wick decomposition procedure\cite{AGD63,Kita-Text} for $\hat{\tilde c}_{{\bf k}}$ within the order of the approximation we adopt.
The result is expressible in terms of Eqs.\ (\ref{rho_k}) and (\ref{F_k}) as
\begin{align}
\langle\Phi|\hat{\tilde c}_{{\bf k}+{\bf q}}^\dagger\hat{\tilde c}_{{\bf k}'-{\bf q}}^\dagger \hat{\tilde c}_{{\bf k}'}\hat{\tilde c}_{{\bf k}}|\Phi\rangle
=&\,
\delta_{{\bf q}{\bf 0}}\rho_{{\bf k}} \rho_{{\bf k}'}+\delta_{{\bf k}',{\bf k}+{\bf q}}\rho_{{\bf k}+{\bf q}} \rho_{{\bf k}}
\notag \\
&\,+
\delta_{{\bf k}',-{\bf k}} F_{{\bf k}+{\bf q}}^* F_{{\bf k}} .
\label{Wick-H_2}
\end{align}

Using Eqs.\ (\ref{c_0-approx}), (\ref{nFW}), and (\ref{Wick-H_2}) in the evaluation of Eq.\ (\ref{calE-def}) and collecting terms proportional to $U_0$, we obtain Eq.\ (\ref{calE}).

\section{Calculation of $A_1({\bf k})$ and $A_2({\bf k})$\label{App:A_12}}

Noting that $\hat{\tilde c}_{{\bf k}} =\hat\beta_{{\bf 0}}^\dagger\hat{c}_{{\bf k}}$,
we express the commutator in Eqs.\ (\ref{A_1}) and (\ref{A_2}) as
\begin{align}
[\hat{\tilde c}_{\bf k},\hat H]=\hat\beta_{{\bf 0}}^\dagger[\hat{c}_{{\bf k}},\hat H]+[\hat\beta_{{\bf 0}}^\dagger,\hat H]\hat{c}_{{\bf k}}.
\label{[c_k,H]}
\end{align}
Subsequently, we substitute Eq.\ (\ref{H-decomp}) into the right-hand side.
The commutator $[\hat\beta_{{\bf 0}}^\dagger,\hat H]$ can be evaluated by using
\begin{align}
\bigl[\hat\beta_{{\bf 0}}^\dagger,(\hat c_{{\bf 0}}^\dagger)^m\hat c_{{\bf 0}}^n\bigr]|\Phi\rangle 
\approx -\frac{m+n}{2}  {\cal N}_{{\bf 0}}^{(m+n)/2-1}(\hat\beta_{{\bf 0}}^\dagger)^{m+1}\hat\beta_{{\bf 0}}^{n}|\Phi\rangle ,
\label{c_0-approx2}
\end{align}
which holds within the same order of approximation as Eq.\  (\ref{c_0-approx}); this may be seen by replacing 
$|\Phi\rangle$ above by $| {\cal N}_{\bf 0}\rangle_{\bf 0}$ of Eq.\ (\ref{|n>_0-def})
and calculating the commutator explicitly to the leading order in ${\cal N}_{\bf 0}$.
On the other hand, $[\hat{c}_{{\bf k}},\hat H]$ in Eq.\ (\ref{[c_k,H]}) can be calculated straightforwardly. 
After that, we can use the procedure for deriving Eq.\ (\ref{nFW}) to evaluate Eqs.\ (\ref{A_1}) and (\ref{A_2}).

We first focus on Eq.\ (\ref{A_1}) and express it as a sum of the four contributions in Eq.\ (\ref{H-decomp}) for convenience.
The results for  $A_{1,\alpha}({\bf k})\equiv \langle\Phi|\bigl[\hat{\tilde c}_{{\bf k}},\hat H_\alpha\bigr]\hat{\tilde c}_{{\bf k}}^\dagger|\Phi\rangle$
($\alpha=0,1,\frac{3}{2},2$) are summarized as follows:
\begin{subequations}
\label{A_1-results}
\begin{align}
A_{1,0}({\bf k})=-\bar n_{{\bf 0}}U_0A_0({\bf k}),
\end{align}
\begin{align}
&\,A_{1,1}({\bf k})
\notag \\
=&\,\left[\varepsilon_k+\bar n_{{\bf 0}}(U_0+U_k)-\frac{1}{{\cal V}}{\sum_{{\bf k}'}}' (U_0+U_{k'})\rho_{{\bf k}'} \right]
\notag \\
&\,\times A_0({\bf k})
+\bar n_{{\bf 0}}U_k F_{{\bf k}} -\frac{1}{{\cal V}}{\sum_{{\bf k}'}}'U_{k'}F_{{\bf k}'} A_0({\bf k}),
\\
&\,A_{1,\frac{3}{2}}({\bf k})
\notag \\
=&\,\frac{\sqrt{{\cal N}_{{\bf 0}}}}{{\cal V}}{\sum_{{\bf k}_2{\bf k}_3}}'\delta_{{\bf k}+{\bf k}_2+{\bf k}_3,{\bf 0}}
w_{{\bf k}{\bf k}_2{\bf k}_3} u_{{\bf k}}u_{{\bf k}_2}u_{{\bf k}_3}
\notag \\
&\,\times \bigl[U_{k_2}(\phi_{{\bf k}}+\phi_{{\bf k}_2}\phi_{{\bf k}_3})
+(U_k+U_{k_2})(\phi_{{\bf k}_3}+\phi_{{\bf k}}\phi_{{\bf k}_2})\bigr]
\notag \\
&\, -\frac{A_0({\bf k})}{{\cal V}\sqrt{{\cal N}_{{\bf 0}}}}{\sum_{{\bf k}_1{\bf k}_2{\bf k}_3}}'\delta_{{\bf k}_1+{\bf k}_2+{\bf k}_3,{\bf 0}}U_{k_1}W_{{\bf k}_1{\bf k}_2;-{\bf k}_3},
\label{A_13/2}
\\
&\,A_{1,2}({\bf k})
\notag \\
=&\,\frac{A_0({\bf k})}{{\cal V}}{\sum_{{\bf k}'}}' (U_0+U_{|{\bf k}-{\bf k}'|})\rho_{{\bf k}'} 
+\frac{F_{\bf k}}{{\cal V}}{\sum_{{\bf k}'}}' U_{|{\bf k}-{\bf k}'|}F_{{\bf k}'} .
\end{align}
\end{subequations}
Substituting Eqs.\ (\ref{A_0}) and (\ref{A_1-results}) into Eq.\ (\ref{barE's1}), we obtain Eq.\ (\ref{barE}) given in terms of Eqs.\ (\ref{xi-def}) and (\ref{Delta-def}).

Calculations of $A_{2,\alpha\alpha'}({\bf k})\equiv \langle\Phi|\bigl[\hat{\tilde c}_{{\bf k}},\hat H_\alpha\bigr]\bigl[\hat H_{\alpha'},\hat{\tilde c}_{{\bf k}}^\dagger]|\Phi\rangle$
($\alpha,\alpha'=0,1,\frac{3}{2},2$) can be performed similarly but rather tediously. Let us write it as
\begin{align}
A_{2,\alpha\alpha'}({\bf k})=\frac{A_{1,\alpha}({\bf k})A_{1,\alpha'}({\bf k})}{A_0({\bf k})}+B_{2,\alpha\alpha'}({\bf k}) .
\label{A_2-decomp}
\end{align}
It then follows that the $B_{2,\alpha\alpha'}({\bf k})$ that contribute to Eq.\ (\ref{barE's2}) are finite only for the combinations of
$(\alpha,\alpha')=(1,1), (1,\frac{3}{2}), (\frac{3}{2},\frac{3}{2})$ up to the leading order in the weak-coupling region with $B_{2,1\frac{3}{2}}({\bf k})=
B_{2,\frac{3}{2}1}({\bf k})$.
Moreover, it is the third term in Eq.\ (\ref{H_1}) that makes  $B_{2,11}({\bf k})$ and $B_{2,1\frac{3}{2}}({\bf k})$ finite. 
Specifically, we obtain
\begin{subequations}
\label{B_2's}
\begin{align}
&\, B_{2,11}({\bf k}) = (\bar n_{{\bf 0}} U_k)^2\left[\rho_{{\bf k}}-\frac{F_{{\bf k}}^2}{A_0({\bf k})}\right],
\end{align}
\begin{align}
&\,B_{2,1\frac{3}{2}}({\bf k})
\notag \\
=&\, \bar n_{{\bf 0}}U_k\frac{\sqrt{{\cal N}_{{\bf 0}}}}{{\cal V}}{\sum_{{\bf k}_2{\bf k}_3}}' 
\delta_{{\bf k}_2+{\bf k}_3+{\bf k},{\bf 0}}w_{{\bf k}{\bf k}_2{\bf k}_3} u_{{\bf k}}u_{{\bf k}_2}u_{{\bf k}_3}
\notag \\
&\,\times \biggl\{U_{k_2} 
\biggl[1+\phi_{{\bf k}}\phi_{{\bf k}_2}\phi_{{\bf k}_3} -\frac{F_{\bf k}}{A_{0}({\bf k})}(\phi_{{\bf k}} +\phi_{{\bf k}_2}\phi_{{\bf k}_3})
\biggr]
\notag \\
&\,
+(U_k+U_{k_2}) \biggl[\phi_{{\bf k}_2}\!+\!\phi_{{\bf k}}\phi_{{\bf k}_3} -\frac{F_{\bf k}}{A_{0}({\bf k})}(\phi_{{\bf k}}\phi_{{\bf k}_2}\!+\!\phi_{{\bf k}_3})\biggr]\biggr\} ,
\end{align} 
\begin{align}
&\,B_{2,\frac{3}{2}\frac{3}{2}}({\bf k})
\notag \\
=&\, \frac{{\cal N}_{{\bf 0}}}{{\cal V}^2}{\sum_{{\bf k}_2{\bf k}_3}}' \delta_{{\bf k}+{\bf k}_2+{\bf k}_3,{\bf 0}}
\bigl[ U_{k_2}(U_{k_2}+U_{k_3})u_{{\bf k}_2}^2u_{{\bf k}_3}^2
\notag \\
&\,
+2(U_{k}+U_{k_2})(U_{k_2}+U_{k_3})F_{{\bf k}_2} u_{{\bf k}_3}^2
\notag \\
&\,
+(U_{k}+ U_{k_2})^2 v_{{\bf k}_2}^2u_{{\bf k}_3}^2
\notag \\
&\,
+(U_{k}+U_{k_2})(U_{k}+U_{k_3}) F_{{\bf k}_2} F_{{\bf k}_3}\bigr] .
\end{align} 
\end{subequations}
Substituting Eqs.\  (\ref{A_2-decomp}) and (\ref{B_2's}) into Eq.\ (\ref{barE's2}) and using Eqs.\ (\ref{nFW}) and  (\ref{A_0}), 
we obtain Eq.\ (\ref{barDE}) up to the leading order.

\end{document}